\documentclass[structabstract]{aa}
\usepackage{graphicx}
\usepackage{lscape}
\usepackage{epsfig}
\usepackage{subfig}
\usepackage{natbib}
\usepackage[figuresright]{rotating}
\bibpunct{(}{)}{;}{a}{}{,}
\usepackage{txfonts}
\begin{document}
\title{Near-IR search for lensed supernovae behind galaxy clusters: \\III. Implications for cluster modeling and cosmology\thanks{Based on observations made with the NASA/ESA Hubble Space Telescope, obtained from the Data Archive at the Space Telescope Science Institute, which is operated by the Association of Universities for Research in Astronomy, Inc., under NASA contract NAS 5-26555. These observations are associated with programs \# 9134, 9289 and 10150.}}
\titlerunning{Near-IR search for lensed supernovae behind galaxy clusters: III.}

\author{T.~Riehm\inst{1,2} 
	\and E.~M\"{o}rtsell\inst{2,3} \and A.~Goobar\inst{2,3} \and R.~Amanullah\inst{2,3} \and T.~Dahl\'en\inst{4} \and J.~J\"{o}nsson\inst{2,3} \and M.~Limousin\inst{5,6}  \and K.~Paech\inst{7} \and J.~Richard\inst{8,6}}

\offprints{T. Riehm}

\institute{
Department of Astronomy, Stockholm University, Albanova University Center, SE 106 91 Stockholm, Sweden \\ 
\email{teresa@astro.su.se} 
\and
The Oskar Klein Centre, Stockholm University, SE 106 91 Stockholm, Sweden 
\and
Department of Physics, Stockholm University, Albanova University Center, SE 106 91 Stockholm, Sweden 
\and
Space Telescope Science Institute, 3700 San Martin Drive, Baltimore, MD 21218, USA
\and
Laboratoire d’Astrophysique de Marseille, Universit\'e de de Provence, CNRS, 38 rue Fr\'ed\'eric Joliot-Curie, F-13388 Marseille Cedex 13, France 3 
\and
Dark Cosmology Centre, Niels Bohr Institute, University of Copenhagen, Juliane Maries Vej 30, DK-2100 Copenhagen, Denmark
\and
Physikalisches Institut Universitat Bonn, Nussallee 12 53115 Bonn, Germany
\and
CRAL, Observatoire de Lyon, Universit\'e Lyon 1, 9 Avenue Ch. Andr\'e, 69561 Saint Genis Laval Cedex, France
}

\date{Received -; accepted -}
\abstract
{Massive galaxy clusters at intermediate redshifts act as
  gravitational lenses that can magnify supernovae (SNe) occurring in background galaxies.} 
  {We assess the possibility to use lensed SNe to put constraints on the mass models of galaxy clusters and the Hubble parameter at high redshift.} 
  {Due to the standard candle nature of Type Ia supernovae (SNe~Ia),
    observational information on the lensing magnification from an
    intervening galaxy cluster can be used to constrain the
    model for the cluster mass distribution. A statistical analysis using parametric cluster
    models was performed to investigate the possible improvements from
    lensed SNe~Ia for the accurately modeled galaxy cluster A1689 and
    the less well constrained cluster A2204. Time delay measurements
    obtained from 
    SNe lensed by accurately modeled galaxy clusters
    can be used to measure the Hubble parameter. 
    For a survey of A1689 we estimate the expected rate of detectable
    SNe~Ia and of multiply imaged SNe.} 
{The velocity dispersion and core radius of the main cluster potential
  show strong correlations with the predicted magnifications and can
  therefore be constrained by observations of SNe~Ia in background galaxies.
  This technique proves especially powerful for galaxy clusters with
  only few known multiple image systems. The main uncertainty for
  measurements of the Hubble parameter from the time delay of strongly
  lensed SNe is due to cluster model uncertainties. 
  For the extremely well modeled cluster A1689, a single time delay measurement could be used to determine
  the Hubble parameter with a precision of $\sim$~10$\%$.} 
{Observations of SNe~Ia behind galaxy clusters can be
  used to improve the mass modeling of the large scale component of
  galaxy clusters and thus the distribution of dark matter.
  Time delays from SNe strongly lensed by accurately modeled galaxy clusters can be used to measure the Hubble constant at high redshifts.} 

\keywords{cosmology: gravitational lensing -- supernovae: general -- galaxies: clusters: general -- galaxies: halos -- dark matter -- cosmological parameters} 


\maketitle


\section{Introduction}\label{sec:intro}
Massive clusters have been successfully employed as gravitational
telescopes (also known as Zwicky telescopes) in order to probe astronomical 
objects at very high
redshifts \citep[e.g.,][]{Kneib04,Bradley08,Bradac09,Zheng09}. More recently, the use of near-IR observations to
look for gravitationally magnified supernovae (SNe) behind massive
galaxy clusters has been investigated. Such data give
insight into the star formation rate at high redshifts, the progenitor
systems of SNe and, if the cluster potential can be estimated
properly, extend the SN~Ia Hubble diagram up to redshift $z \sim 3$ or possibly
even higher. Such measurements would dramatically extend the redshift baseline for which the expansion history of the universe can be probed, which could turn out to be essential for understanding the nature of dark energy.

A first transient object behind the massive cluster A1689 was found in
a ground based pilot survey at ESO using the ISAAC \citep{moorwood98} 
camera on VLT
\citep[][ hereafter Paper I and Paper II, respectively]{Stanishev09,Goobar09}. The transient was consistent with a reddened Type IIP SN at $z$ = 0.59 with a lensing
magnification $\Delta m$~=~1.4 mag. Other SN candidates have been found in
a survey with the newer HAWK-I \citep{pirard04,casali06,kissler08}
 camera at VLT \citep{Amanullah11}. Although not optimally cadenced for the purpose, 
HST detections of lensed SNe may also result from
CLASH \citep{Postman11}, part of the Multi-Cycle Treasury (MCT) program targeting 25 massive
cluster, both at optical and near-IR wavelengths. 

In this paper we investigate the impact observations of 
gravitationally lensed SNe
could have on the modeling of galaxy cluster potentials. We will
focus on the detection feasibility with either an 8-m class NIR
survey like the one with HAWK-I (FOV $\approx 56$ arcmin$^2$), or with the 
Wide Field Camera 3 instrument on board the Hubble Space Telescope (HST/WFC3) with a 10
times smaller FOV. 
Since SNe~Ia have small intrinsic luminosity dispersion (after correction for lightcurve shape and reddening), we can estimate the absolute magnification of the SNe
and thus break the so called mass sheet degeneracy of gravitational
lenses. This degeneracy implies that the density distribution of the lens 
can always be rescaled and a constant-density mass-sheet added such that the, 
also properly rescaled, source plane is projected onto the same observed 
images. Thereby, by breaking this degeneracy, direct constraints on the distribution of the dark matter component in galaxy clusters can be obtained.
 
If a SN is strongly lensed, a measurement of the time delay between the 
transient in the multiple images could
potentially constrain the Hubble parameter \citep{Refsdal64}, and
thus dark matter and dark energy parameters \citep{Goobar02,Mortsell06,Suyu10}. Though not as precise as other
cosmological tests to study dark energy, the time delay technique has
the major advantage of measuring cosmological parameters at redshifts
where few other probes are currently available.  Given the transient
nature of SNe, the time delay between multiple images could
potentially be measured to very high precision.  The main limitation
of this technique is the strong degeneracy between the lens mass model
and the Hubble parameter, $H_0$ 
\citep[e.g.,][]{Wambsganss94,Witt00,Kochanek02,Zhao03}. However, observing
cluster lenses with potentials well constrained by a large number of
already known multiple images or, in the case of a strongly lensed SNe~Ia, 
direct magnification information, this degeneracy can be
diminished \citep{Oguri03}.  This technique is applicable to any
object with a variable light curve which would make it possible to
determine a time delay between different images, including quasars
(QSOs) and gamma-ray bursts (GRBs).  Although, there are $\sim$~100
strongly lensed QSOs currently known, time delays have been measured
for only $\sim$~20 of them \citep{Oguri07}. This is due to the fact that most
QSOs show variability only at a fairly low level, making time delay
determinations challenging. Time delay measurements of
QSOs behind cluster lenses have so far only been possible in two
cases: SDSSJ1029+2623 \citep[three images,][]{Inada06} and SDSS J1004+4112 \citep[five images,][]{Fohlmeister08}. 
Image magnifications of
about $\Delta m \sim$ 4.5 mag have been recorded. However, the mass
distributions in these clusters are not modeled well enough to
constrain the Hubble constant.  Although the time delay of a GRB,
could be determined with very high precision, they are expected to be
too rare to admit a survey of lensed GRBs behind
clusters. Therefore, in this paper we will focus on time delays for
multiply imaged SNe.

The paper is structured as follows: In Sect.~\ref{sec:clustermod},
we describe the cluster model used in our analysis.
Sect.~\ref{sec:SNe} 
investigates the observational prospects for lensed SNe
behind a cluster. Constraints on the cluster modeling and the Hubble
constant from lensed SNe are studied in Sects.~\ref{sec:modelconstr}
and \ref{sec:hubble}. Finally, we discuss our results and conclude in
Sect.~\ref{sec:dis}.

Throughout the paper, we assume the cosmological parameters to
be $\Omega_{\rm M}=0.3$, $\Omega_{\Lambda}=0.7$, and $H_0= 70$ km s$^{-1}$ 
Mpc$^{-1}$.


\section{Cluster modeling}\label{sec:clustermod}

\begin{figure}[t]
 \caption{
  Magnification map of A1689 overlaid on top of an HST ACS image of
  the cluster. The $\Delta m =1,2,3$ mag (pink, blue, magenta)
  contours are shown for a source at $z=2$. The green (orange) circles
  indicate the positions of strongly lensed galaxies with
  spectroscopic redshifts and time delays below (above) 5 years.  Red
  points indicate strongly lensed sources with photometric redshifts.}
  \label{fig:systems} 
  \begin{center} 
    \includegraphics[width=9cm]{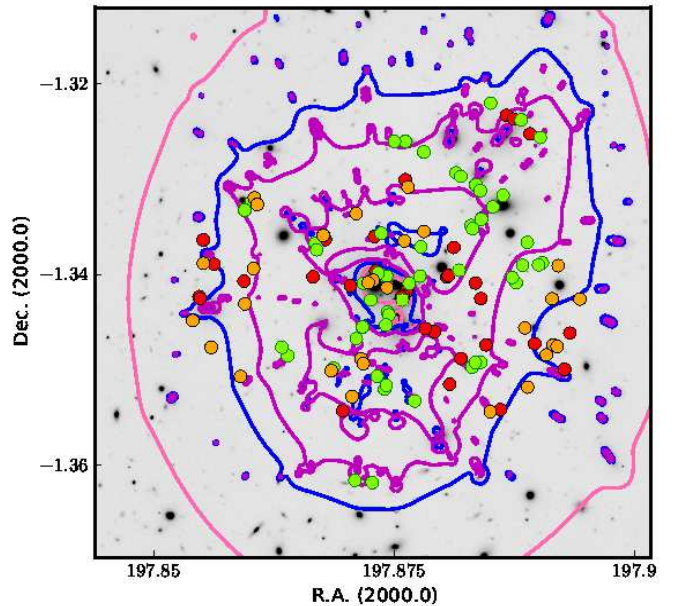}
  \end{center}
\end{figure}

The focus of this paper is the massive cluster
A1689 at redshift $z = 0.187$ which is one of the best studied
clusters. For this cluster there are  34 known multiply imaged
background galaxies and a total
of 114 images. For 24 of the 34 background systems, there are secure
spectroscopic redshifts available, ranging from $z$ = 1.1 to 4.9
\citep{Broadhurst05,Limousin07} including a total of 82
images (see Fig.~\ref{fig:systems}). For this cluster, a detailed mass
model is available consisting of 272 parametrized
pseudo-isothermal elliptical mass distributions (PIEMD). 
This is an updated version of
the mass model presented in \citet{Limousin07}. Each potential is
characterized using seven parameters: the center position (RA, DEC),
the position angle $\theta$, and the parameters of the mass profile:
the central velocity dispersion $\sigma$, the ellipticity $e$, the
core radius $r_{\rm core}$ and the cut radius $r_{\rm cut}$. The model has 33
free parameters describing the mass distribution of the two large
scale dark matter clumps (Clump 1 and Clump 2), the dark matter
halos of three individual galaxies which were found to play an
essential role for producing multiply imaged systems (BCG, Galaxy
1 and Galaxy 2), as well as two scaling relations based on luminosity
for the remaining identified cluster galaxies, $L^*r_{\rm cut}$ and
$L^*\sigma$ (compare Table~\ref{tab:par}). 
The latter parameters are used to assign masses to 
the cluster galaxies, which are mainly early-type galaxies, via the
scaling relations, $r_{\rm cut}(L) = r_{\rm cut}(L/L^*)^{1/2}$ and $\sigma(L) = \sigma(L/L^*)^{1/4}$, where $L^*$ is the luminosity of a typical galaxy in the cluster.

\begin{table*}
\caption{Parameter values inferred for the dark matter clumps considered in the optimization procedure for galaxy cluster A1689. Coordinates are given in arcseconds with respect to the BGC. Error bars correspond to 1$\sigma$ confidence level as inferred from the MCMC optimization. The model has a total of 33 free parameters. Values in brackets are not optimized.}
\label{tab:par}
\centering
\begin{tabular}{l c c c c c c c}
 \hline \hline 
Clump & RA & DEC & $e$ & $\theta$ & $r_{\rm core}$ (kpc) & $r_{\rm cut}$ (kpc) & $\sigma$ (km s$^{-1}$) \\
\hline 
Clump 1 & $   0.5^{+ 0.2}_{ -0.2}$ & $  -8.5^{+ 0.4}_{ -0.4}$ & $  0.21^{+0.01}_{-0.01}$ & $  90.4^{+ 1.2}_{ -1.1}$ & $ 101.2^{+4.6}_{-4.0}$ & [1515.7] & $ 1445.5^{+16.0}_{-14.7}$ \\
Clump 2 & $ -70.7^{+ 1.3}_{ -1.6}$ & $  49.1^{+ 3.0}_{ -3.6}$ & $  0.77^{+0.04}_{-0.05}$ & $  78.6^{+ 2.6}_{ -2.6}$ & $ 67.9^{+8.0}_{-5.8}$ & [501.0] & $ 647.1^{+ 2.1}_{ -4.9}$ \\  
BCG & $  -1.1^{+ 0.3}_{ -0.2}$ & $   0.1^{+ 0.4}_{ -0.4}$ & $  0.48^{+0.04}_{-0.04}$ &  $  66.0^{+ 6.9}_{ -6.3}$ & $  5.3^{+1.1}_{-1.1}$ & $ 130.8^{+37.2}_{-36.2}$ & $ 441.7^{+11.6}_{-12.2}$ \\  
Galaxy 1 & [49.1] & [31.5] & $  0.70^{+0.10}_{-0.15}$ &  $ 113.6^{+ 7.2}_{ -9.8}$ & $ 25.6^{+3.0}_{-3.4}$ & $ 161.2^{+13.3}_{-21.9}$ & $ 262.8^{+12.1}_{-16.6}$  \\  
Galaxy 2 & $ -45.7^{+ 0.4}_{ -0.7}$ & $  31.6^{+ 0.8}_{ -0.9}$ & $  0.80^{+0.05}_{-0.04}$ &   $  45.5^{+ 2.1}_{ -2.1}$ & $ 17.6^{+1.8}_{-3.0}$ & $ 182.8^{+5.2}_{-9.7}$ & $ 425.6^{+21.5}_{-28.4}$ \\  
L$^*$ elliptical galaxy & \ldots & \ldots & \ldots & \ldots & [0.15] &
$ 54.5^{+3.7}_{-6.7}$ & $ 154.6^{+ 5.3}_{ -3.4}$ \\
\hline
\end{tabular}
\end{table*}

\begin{table*}
\caption{Parameter values inferred for the dark matter clump and the 
scaling relation considered in the optimization procedure for galaxy
cluster A2204. Coordinates are given in arcseconds with respect to the
BGC. Error bars correspond to 1$\sigma$ confidence level as inferred
from the MCMC optimization. The model has a total of 5 free
parameters. Values in brackets are not optimized.}
\label{tab:A2204par}
\centering
\begin{tabular}[t]{l c c c c c c c}
\hline \hline
Clump & RA & DEC & $e$ & $\theta$ & $r_{\rm core}$ (kpc) & $r_{\rm cut}$ (kpc) & $\sigma$ (km s$^{-1}$) \\
\hline 
Clump 1 & [0.0] & [0.0] & $  0.38^{+0.18}_{0.14}$ & $ 127.64^{+5.07}_{6.63}$ & $ 54.64^{+18.05}_{22.34}$  & [1000.0] &  $ 933.47^{+160.66}_{167.06}$\\ 
L$^*$ elliptical galaxy & \ldots & \ldots & \ldots & \ldots & [0.15] &
[45] & $ 203.30^{+24.05}_{28.94}$ \\
\hline
\end{tabular}
\end{table*}

For investigating the power of lensed Type Ia SNe observations for constraining 
cluster models we are also studying cluster A2204 at redshift $z$ = 0.1524. 
For this cluster only one
multiply lensed image system at $z = 1.06$ is known, making it much
less constrained than A1689. The
mass model for this cluster consists of one main potential, describing
the smooth cluster potential, and 34 cluster member galaxy potentials. 
All potentials are described by PIEMD profiles and the model  
has a total of 5 free parameters: the ellipticity, $e$, position
angle, $\theta$, core radius, $r_{\rm core}$, and velocity dispersion,
$\sigma$, of the main potential as well as the mass-luminosity scaling
relation for the cluster galaxies, $L^{*}\sigma$ 
\citep[compare Table~\ref{tab:A2204par}, for details see][]{Richard10}.

{\tt LENSTOOL} is a software package for modeling mass
distributions of galaxies and clusters in the strong and weak lensing
regime \citep[][ http://www.oamp.fr/cosmology/lenstool/]{Jullo07}.
Monte Carlo Markov Chains (MCMC) are used to constrain the free parameters 
of the cluster model derived from observational data of the
background galaxies. The output contains a chain sampling the probability
distribution of parameter values, each of which corresponds to a
specific model {\em realization}. The density of parameter values
gives a measure of the corresponding probability distribution for the
parameters and can be used to estimate the errors of the corresponding parameters. 
For each of the realizations, the magnification and time delay function at any 
given position behind the cluster, can be computed.

Assuming that we observe the magnification or time delay of a multiply
imaged SN, we are able to rule out all realizations that do not agree
with these (simulated) additional constraints. By comparing the
remaining realizations to the original chain, we are then able to
judge how powerful the additional constraints are in constraining the
cluster potential.


\section{Supernovae behind clusters}\label{sec:SNe}

To investigate how efficient massive galaxy clusters are 
for detecting distant SNe and putting constraints on cluster
modeling or cosmology, a good estimate on how many SNe behind the
cluster that can be expected during a given survey time is needed. For
example, for a 5 year monthly survey of one very massive A1689-like
cluster with HAWK-I, it has been shown that the total number of SNe
expected in the background galaxies is on the order of 40--70 SNe, out
of which approximately 20--30 are SNe~Ia (depending on the
underlying rates estimate for the various SN types, see Paper II).

For the purpose of studying time delays using galaxy cluster A1689, we
focus primarily on potential SN explosions in the 24 known multiply
lensed background galaxies with spectroscopic redshifts. Ten 
multiply imaged galaxies with photometric redshifts only are not taken into 
consideration to avoid additional sources of error due to uncertainties 
in photometric redshifts. Furthermore, with
more data being collected, new multiple image systems might be
detected. The numbers we present below should therefore be considered
as lower limits.

Since the rate of SNe is expected to be coupled to the star-formation
rate (SFR) in these galaxies, we use rate predictions for the strongly
lensed systems derived from local estimates of the SFR. As the UV
luminosity is dominated by the most short-lived stars, it is closely
related to star formation. Thus, we use $L_{2800}$, the flux at
rest-frame $\lambda_{\rm eff} = 2800$ \AA \ (redshifted to optical bands)
as a tracer of the SFR in the strongly lensed galaxies. Since there
are very deep HST observations of A1689 with multiple filters, these
estimates are remarkably precise.

The absolute $L_{2800}$ magnitude is derived after taking into account
the distance modulus, K-corrections, extinction and lensing
magnification from the cluster model. Finally, we use the relation
between $L_{2800}$ and SFR from \citet{Dahlen07} to relate the flux to
star formation,
\begin{equation}
{\rm SFR} \left(M_\odot {\rm yr}^{-1}\right) = L_{2800} \left({\rm erg\cdot s}^{-1}
{\rm Hz}^{-1}\right) \times \left(7.0 \cdot 10^{27}\right)^{-1}\, .
\end{equation}
The expected rate for core collapse SNe is calculated from 
\begin{equation}
r_{\rm CC} (z) = k^{50}_8 \times {\rm SFR}(z)\, ,
\end{equation}
where $k^{50}_8 = 0.007 M^{-1}_{\odot}$ is estimated using a Salpeter
IMF and a progenitor mass range of between 8 and 50 solar masses.  For
SNe~Ia, we assume that the rate has two components, one proportional to
the SFR and one proportional to stellar mass, according to the model
of \citet{Scannapieco05}. The rate is given by
\begin{equation}
r_{\rm Ia}(t)= A \cdot {\rm SFR}(t) + B \cdot M(t)\, .
\end{equation} 
The mass, $M$, of individual galaxies are calculated using mass-to-light
ratios from \citet{Bell03}. The resulting rates for core collapse SNe
and SNe~Ia in multiply lensed background galaxies with
spectroscopic redshifts and predicted time delays below 5 years are
summarized in Table~\ref{tab:spec_delays}. In total, we expect
$\sim$~0.6 core collapse SNe and $\sim$~0.05 SNe~Ia per year to
explode in the known multiply imaged galaxies with spectroscopic
redshifts.

\begin{table*}
 \caption{All image pairs of multiply lensed background galaxies with
  spectroscopic redshift and a predicted time delay of less than 5
  years. Column 3 gives the predicted time delay between the images. A
  negative time-delay indicates that image 2 will occur before image
  1. Columns 4 and 5 give the predicted magnification for image 1 and
  image 2, respectively. In column 6 and 7 our rate estimates are
  shown for core collapse SNe and SNe~Ia,
  respectively. Columns 8--14 indicate the detectability of a lensed
  SN for the different SN types assuming a limiting magnitude of 25.25 mag
  (Vega) in F140W, e.g. in a survey using 1/3 orbit with the HST/WFC3 instrument
  (compare with Fig.~\ref{fig:SN_lc}). Errors are given as one sigma.}
  \label{tab:spec_delays}
  \centering
  \begin{tabular}{r r r r r r r c c c c c c c}
    \hline \hline
     Images & $z_{\rm spec}$ & \multicolumn{1}{c}{$\Delta t$} & \multicolumn{1}{c}{$\Delta m_1$} & \multicolumn{1}{c}{$\Delta m_2$} & \multicolumn{1}{c}{$r_{\rm CC}$ $\times$  $10^{2}$} & \multicolumn{1}{c}{$r_{\rm Ia}$ $\times$ $10^{3}$} &  \multicolumn{7}{c}{Detectable SN types} \\
 & & \multicolumn{1}{c}{(d)} & \multicolumn{2}{c}{(mag)} &  \multicolumn{2}{c}{(yr$^{-1}$)} & Ia & IIP & IIL & IIL$_{\rm b}$ & IIn & Ib/c & HN \\
     \hline
  1.1 +   1.2 &  3.04 & $-83.2^{+76.9}_{-69.4}$ & $  3.80^{+0.30}_{-0.29}$ & $ 5.58^{+1.21}_{-0.74}$ & $  2.07^{+  1.07}_{  1.04}$ & $   1.32^{+  0.68}_{  0.66}$ &  y  &  y  &  y  &  y  &  y  &  y  &  y  \\
  1.3 +   1.6 &  3.04 & $1248.2^{+135.5}_{-148.9}$ & $
 2.12^{+0.05}_{-0.05}$ & $ 0.88^{+0.08}_{-0.07}$ & \ldots  & \ldots &  y  &    &    &    &  y  &    &   \\
  1.4 +   1.5 &  3.04 & $-139.4^{+87.7}_{-79.7}$ & $
 2.85^{+0.07}_{-0.06}$ & $ 1.88^{+0.05}_{-0.04}$ & \ldots & \ldots & y  &    &    &  y  &  y  &    &  y  \\
  2.2 +   2.3 &  2.53 & $-163.0^{+77.7}_{-72.2}$ & $  2.97^{+0.07}_{-0.06}$ & $ 2.01^{+0.05}_{-0.04}$ & $  1.74^{+  0.85}_{  0.85}$ & $1.12^{+  0.52}_{  0.53}$   &  y  &    &  y  &  y  &  y  &  y  &  y   \\
  2.4 +   2.5 &  2.53 & $1072.8^{+119.4}_{-132.0}$ & $
 2.07^{+0.05}_{-0.05}$ & $ 1.00^{+0.07}_{-0.07}$ & \ldots & \ldots  & y  &    &    &  y  &  y  &    &  y  \\
  4.1 +   4.2 &  1.16 & $575.6^{+63.7}_{-64.9}$ & $  2.96^{+0.04}_{-0.04}$ & $ 2.85^{+0.05}_{-0.05}$ & $  0.44^{+  0.16}_{  0.16}$ & $   0.28^{+  0.10}_{  0.10}$  &  y  &  y  &  y  &  y  &  y  &  y  &  y \\
  5.1 +   5.2 &  2.64 & $-31.2^{+17.6}_{-16.6}$ & $  3.54^{+0.08}_{-0.08}$ & $ 2.82^{+0.07}_{-0.07}$ & $  0.97^{+  0.43}_{  0.43}$ & $   0.65^{+  0.27}_{  0.27}$ &  y  &  y  &  y  &  y  &  y  &  y  &  y  \\
  6.1 +   6.2 &  1.15 & $ 70.8^{+166.5}_{-170.4}$ & $  2.81^{+0.04}_{-0.04}$ & $ 2.95^{+0.12}_{-0.09}$ & $  0.45^{+  0.21}_{  0.21}$ & $   6.51^{+  5.50}_{  5.50}$ &  y  &  y  &  y  &  y  &  y  &  y  &  y \\
  6.1 +   6.3 &  1.15 & $1759.3^{+237.3}_{-249.5}$ & $  2.81^{+0.04}_{-0.04}$ & $ 3.69^{+0.74}_{-0.44}$ & \ldots  & \ldots & y  &  y  &  y  &  y  &  y  &  y  &  y \\
  6.1 +   6.4 &  1.15 & $222.8^{+145.5}_{-161.3}$ & $
 2.81^{+0.04}_{-0.04}$ & $ 3.30^{+0.56}_{-0.33}$ &  \ldots & \ldots &  y  &  y  &  y  &  y  &  y  &  y  &  y  \\
  6.2 +   6.3 &  1.15 & $1680.8^{+262.3}_{-250.1}$ & $
 2.96^{+0.11}_{-0.10}$ & $ 3.69^{+0.74}_{-0.44}$ &  \ldots & \ldots & y  &  y  &  y  &  y  &  y  &  y  &  y \\
  6.2 +   6.4 &  1.15 & $157.3^{+113.0}_{-119.9}$ & $
 2.95^{+0.11}_{-0.10}$ & $ 3.30^{+0.56}_{-0.33}$ & \ldots  & \ldots &  y  &  y  &  y  &  y  &  y  &  y  &  y \\
  6.3 +   6.4 &  1.15 & $-1526.1^{+281.8}_{-286.3}$ & $
 3.73^{+0.73}_{-0.44}$ & $ 3.31^{+0.56}_{-0.32}$ & \ldots & \ldots  &  y  &  y  &  y  &  y  &  y  &  y  &  y\\
  7.2 +   7.3 &  4.87 & $645.0^{+96.0}_{-82.3}$ & $  1.77^{+0.06}_{-0.07}$ & $ 0.31^{+0.13}_{-0.13}$ & $ 13.04^{+  6.06}_{  6.08}$ & $   8.34^{+  3.70}_{  3.71}$  &    &    &    &    &    &    &  \\
 10.2 +  10.3 &  1.83 & $1376.2^{+143.0}_{-148.6}$ & $  2.49^{+0.08}_{-0.08}$ & $ 0.16^{+0.12}_{-0.12}$ &  $  5.42^{+  2.48}_{  2.47}$ & $   4.31^{+  1.58}_{  1.57}$&  y  &    &    &  y  &  y  &    &  y  \\
 12.2 +  12.3 &  1.83 & $ 63.1^{+13.0}_{-12.9}$ & $  3.69^{+0.06}_{-0.06}$ & $ 7.03^{+1.05}_{-0.61}$ & $  0.68^{+  0.32}_{  0.32}$ & $   0.42^{+  0.20}_{  0.20}$ &  y  &  y  &  y  &  y  &  y  &  y  &  y  \\
 14.1 +  14.2 &  3.40 & $-236.8^{+14.1}_{-17.2}$ & $  4.76^{+0.39}_{-0.25}$ & $ 2.92^{+0.06}_{-0.06}$ & \ldots & \ldots &  y  &    &    &  y  &  y  &    &  y\\
 15.1 +  15.3 &  1.82 & $888.6^{+138.8}_{-139.3}$ & $  2.61^{+0.08}_{-0.08}$ & $ 0.74^{+0.14}_{-0.13}$ &  $  0.91^{+  0.42}_{  0.42}$ & $   0.56^{+  0.26}_{  0.26}$  &  y  &    &    &  y  &  y  &    &  y  \\
 17.1 +  17.2 &  2.66 & $-78.8^{+24.6}_{-20.8}$ & $  3.21^{+0.14}_{-0.15}$ & $ 1.40^{+0.09}_{-0.09}$ & $ 15.13^{+  6.21}_{  6.19}$ & $  10.92^{+  3.84}_{  3.82}$  & y  &    &    &  y  &  y  &    &  y  \\
 19.1 +  19.5 &  2.60 & $1507.0^{+89.5}_{-92.0}$ & $  2.40^{+0.06}_{-0.06}$ & $ 1.02^{+0.11}_{-0.10}$ & $  3.58^{+  1.53}_{  1.53}$ & $   2.43^{+  0.96}_{  0.96}$  & y  &    &    &  y  &  y  &    &  y \\
 19.3 +  19.4 &  2.60 & $-84.1^{+10.5}_{-12.7}$ & $
 3.11^{+0.06}_{-0.09}$ & $ 5.14^{+0.21}_{-0.18}$ &  \ldots & \ldots & y  &  y  &  y  &  y  &  y  &  y  &  y \\
 22.1 +  22.2 &  1.70 & $462.2^{+111.6}_{-98.7}$ & $  2.21^{+0.07}_{-0.05}$ & $ 0.13^{+0.22}_{-0.20}$ & $  2.77^{+  1.03}_{  1.06}$ & $   2.16^{+  0.68}_{  0.70}$   & y  &    &    &  y  &  y  &    &  y  \\
 24.2 +  24.3 &  2.63 & $-1029.2^{+157.2}_{-148.9}$ & $  4.81^{+1.14}_{-0.61}$ & $ 2.96^{+0.05}_{-0.05}$ & $  0.93^{+  0.43}_{  0.42}$ & $   0.62^{+  0.27}_{  0.26}$  &  y  &  y  &  y  &  y  &  y  &  y  &  y\\
 24.2 +  24.4 &  2.63 & $-1364.2^{+198.6}_{-196.9}$ & $
 4.81^{+1.14}_{-0.61}$ & $ 2.83^{+0.09}_{-0.08}$ & \ldots & \ldots &  y  &  y  &  y  &  y  &  y  &  y  &  y\\
 24.3 +  24.4 &  2.63 & $-343.0^{+169.7}_{-164.7}$ & $
 2.96^{+0.05}_{-0.05}$ & $ 2.83^{+0.09}_{-0.08}$ &  \ldots & \ldots & y  &  y  &  y  &  y  &  y  &  y  &  y  \\
 29.2 +  29.3 &  2.57 & $1017.8^{+162.5}_{-170.9}$ & $  3.04^{+0.05}_{-0.05}$ & $ 4.90^{+1.27}_{-0.66}$ & $  0.71^{+  0.40}_{  0.37}$ & $   0.53^{+  0.27}_{  0.26}$  & y  &  y  &  y  &  y  &  y  &  y  &  y \\
 29.2 +  29.4 &  2.57 & $136.7^{+147.5}_{-140.0}$ & $  3.04^{+0.05}_{-0.05}$ & $ 2.78^{+0.08}_{-0.08}$ & \ldots & \ldots& y  &  y  &  y  &  y  &  y  &  y  &  y \\
 29.3 +  29.4 &  2.57 & $-881.6^{+179.0}_{-168.9}$ & $  4.90^{+1.27}_{-0.66}$ & $ 2.78^{+0.08}_{-0.07}$ & \ldots  & \ldots & y  &  y  &  y  &  y  &  y  &  y  &  y  \\
 30.1 +  30.2 &  3.05 & $-27.3^{+186.6}_{-191.8}$ & $  2.71^{+0.09}_{-0.07}$ & $ 4.89^{+0.99}_{-0.52}$ & $  3.41^{+  2.48}_{  2.44}$ & $   2.40^{+  1.68}_{  1.65}$  & y  &    &    &  y  &  y  &    &  y \\
 30.1 +  30.3 &  3.05 & $1022.6^{+339.4}_{-316.4}$ & $
 2.71^{+0.09}_{-0.08}$ & $ 4.86^{+1.10}_{-0.70}$ & \ldots  & \ldots & y  &    &    &  y  &  y  &    &  y   \\
 30.2 +  30.3 &  3.05 & $1052.7^{+183.1}_{-166.8}$ & $
 4.89^{+0.98}_{-0.53}$ & $ 4.86^{+1.09}_{-0.70}$ &  \ldots & \ldots & y  &  y  &  y  &  y  &  y  &  y  &  y \\
 32.1 +  32.2 &  3.00 & $350.1^{+90.7}_{-91.6}$ & $  2.15^{+0.05}_{-0.06}$ & $ 2.97^{+0.07}_{-0.07}$ & $  4.38^{+  1.75}_{  1.75}$ & $   3.07^{+  1.07}_{  1.07}$ &  y  &    &    &  y  &  y  &    &  y \\
 32.3 +  32.4 &  3.00 & $759.0^{+98.3}_{-98.9}$ & $
 2.19^{+0.06}_{-0.05}$ & $ 1.29^{+0.06}_{-0.05}$ & \ldots & \ldots  &  y  &    &    &  y  &  y  &    &   \\
 35.1 +  35.3 &  1.91 & $1246.8^{+95.3}_{-95.9}$ & $  2.69^{+0.06}_{-0.06}$ & $ 0.97^{+0.15}_{-0.14}$ & $  2.54^{+  1.07}_{  1.07}$ & $   1.65^{+  0.67}_{  0.67}$ &  y  &    &    &  y  &  y  &    &  y \\
 36.1 +  36.2 &  3.01 & $-45.4^{+18.2}_{-21.0}$ & $
     4.92^{+0.17}_{-0.16}$ & $ 4.52^{+0.13}_{-0.14}$ & $  0.06^{+
     0.03}_{  0.03}$ & $   0.04^{+  0.02}_{  0.02}$  &  y  &  y  &  y
     &  y  &  y  &  y  &  y  \\
\hline
  \end{tabular} 
\end{table*}

To calculate the properties of the lens systems, i.e., masses and SFRs, 
we use available observations in the HST/ACS optical filters (f435w, f475w, 
f555w, f625w, f775w, f850lp), as well as HST/NICMOS IR filters (f110w, f160w) 
from Richard et al. (in preparation).

\begin{figure*}[t]
 \caption{Simulated light curves of SNe~Ia and IIP for pairs of
  images of strongly lensed galaxies behind A1689. Only pairs with
  time-delay $<$ 5 years are shown. Additional information on
  different image pairs labeled by, e.g., 1.2 and 1.1 in the top right
  corner of each sub panel, can be found in
  Table~\ref{tab:spec_delays}. The dashed line indicates an expected
  magnitude limit 25.25 (Vega) in F140W.}
  \label{fig:SN_lc}
  \begin{center}
  \includegraphics[width=16cm]{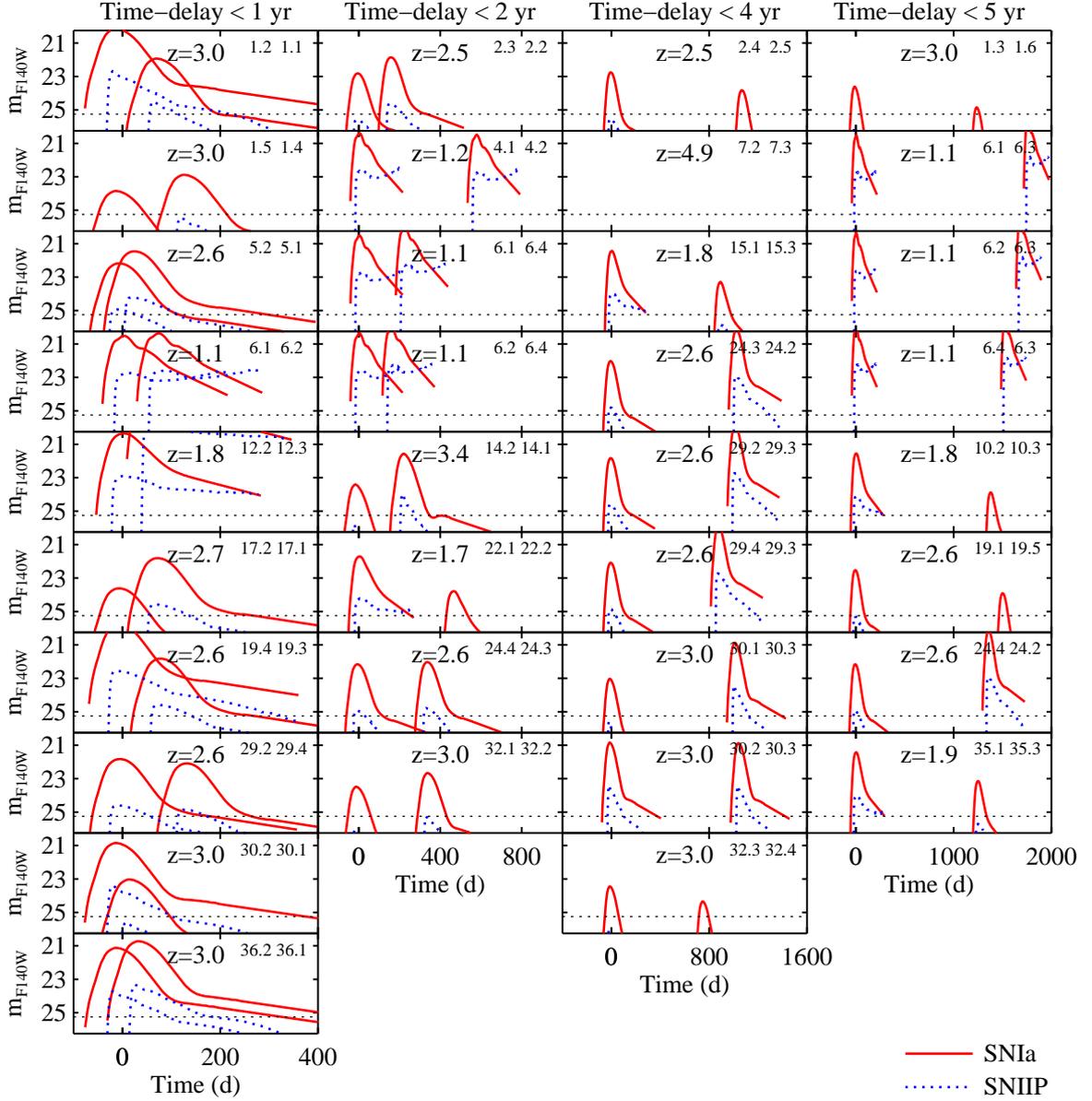}
 
  \end{center}
\end{figure*}

Using the cluster model to calculate magnifications, we can now
investigate the observability of such systems. In order to be
observable, not only must the time delay be smaller than the survey
length (here assumed to be 5 years) but also the luminosity of both
images must lie above the observation threshold. Here we have chosen a
magnitude limit of 25.25 mag (Vega) in F140W, corresponding to what could
be achieved, e.g., with 1/3 orbit with the HST/WFC3. 
In Fig.~\ref{fig:SN_lc}, the
predicted light curves for SNe of Type Ia and IIP are shown. 
As can be seen, nearly all SNe~Ia exploding in one of
the known multiply lensed background galaxies would be bright enough to
be observable, while about half of the SNe IIP, mostly at lower
redshift and with high magnification, would lie above the luminosity
threshold. Columns 8--14 in Table~\ref{tab:spec_delays} indicate
which SN type would be observable for each image combination producing
appropriate time delays.

\section{Cluster model constraints}\label{sec:modelconstr}

When modeling the potential of a cluster as described in
Sect.~\ref{sec:clustermod}, information on the position of the
multiple images is used to optimize the free parameters. In general,
the addition of a constant surface density $\kappa_{\rm c}$ to a lens
potential leaves the resulting image positions of a system
unchanged. This is known as {\em the mass sheet degeneracy}. In the
case of A1689, due to its large number of known multiply imaged
systems at different redshifts, this degeneracy is broken. However,
for galaxy clusters with only very few observed lens systems, this
degeneracy can be problematic. The detection of a SN~Ia behind
a cluster offers a direct way of constraining the lens potential. Due
to the standard candle nature of SNe~Ia, see \citet{GoobarLeibundgut}
for a recent review, the information on the
absolute magnitude can be used to measure the magnification at the
position of the SN and compare it to mass model predictions. One would
then be able to rule out realizations producing deviant
magnifications. The precision of the magnification measurements are
ultimately limited by the intrinsic scatter in the brightness of SNe~Ia
after corrections for lightcurve shape and color, about 0.1 mag 
in the rest-frame optical wavelength region \citep{Conley11}. We do not
anticipate any additional sources of error due to the cluster in the 
line of sight. E.g., observing at near-IR wavelengths ensures that corrections
from dust in the (low-$z$) cluster are small, especially since galaxy clusters
are relatively dust-free environments, see \citet{Dawson09} and references
therein.
Thus, a SN~Ia exploding in any of the background
galaxies behind the cluster could be used to put constraints on the
lensing potential. As discussed in Paper II, we expect to detect
$\sim$~20--30 SNe~Ia (depending on the underlying rates
estimates) to be detectable, e.g., in a 5 year monthly survey at
VLT. In order to assess the power of this method, we investigate the
strength of the correlation between the optimized model parameters and
the magnification for different positions behind the cluster.

\begin{figure*}
\caption{Correlations between the predicted magnification 
as a function of image position for a source at redshift $z = 2$
and the input parameters which are optimized in the model for A1689
(compare Table~\ref{tab:par}) in a field of view $\pm$ 100
arcseconds around the cluster center. Correlations are given as the
absolute values of the correlation coefficients. A high value
(close to 1; dark red) indicates that that the magnification for an
image at the given position will display large variations when the
input parameter is varied. Being able to measure the absolute
magification at this image position, using e.g. SNe~Ia observations,
will thus make it possible to improve the constraints on the cluster
parameter. If the correlation is low (close to 0; light yellow), we do
not expect magnification information to significantly improve the
current constraint on the cluster parameter at hand. The strongest
correlations and best possibility of improving the model once a SN~Ia
is observed, are the ellipticity, $e$, core radius, $r_{\rm core}$,
and velocity dispersion, $\sigma$, of the main lensing potential.
The last panel shows an ACS image of the cluster overlaid with the
critical lines for $z = 2$. The FOV in all panels is identical with
Fig.~\ref{fig:systems}.}
\label{fig:z2correls}
\centering
\begin{tabular}{ccc}
\epsfig{file=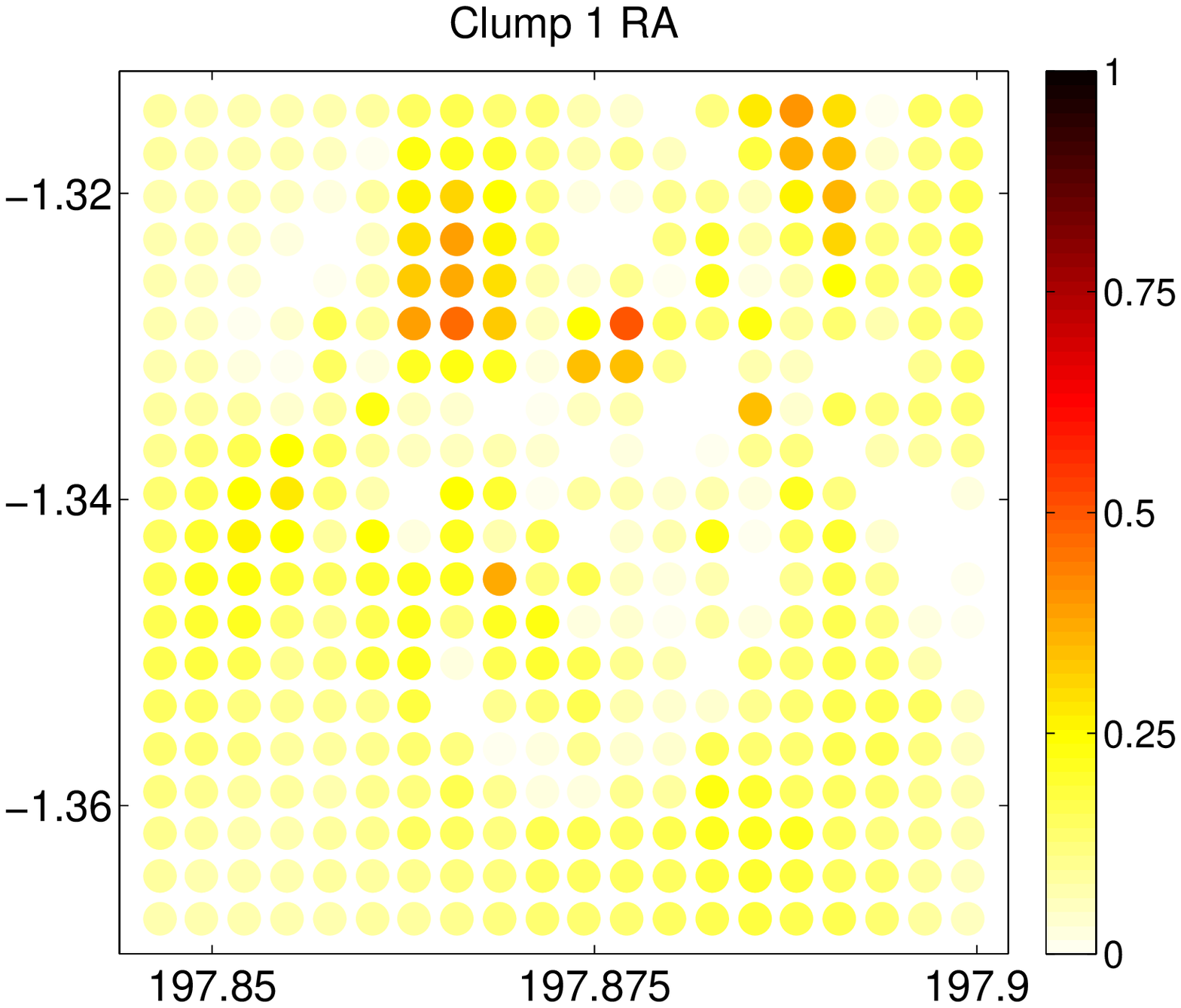,width=0.26\linewidth,clip=} & 
\epsfig{file=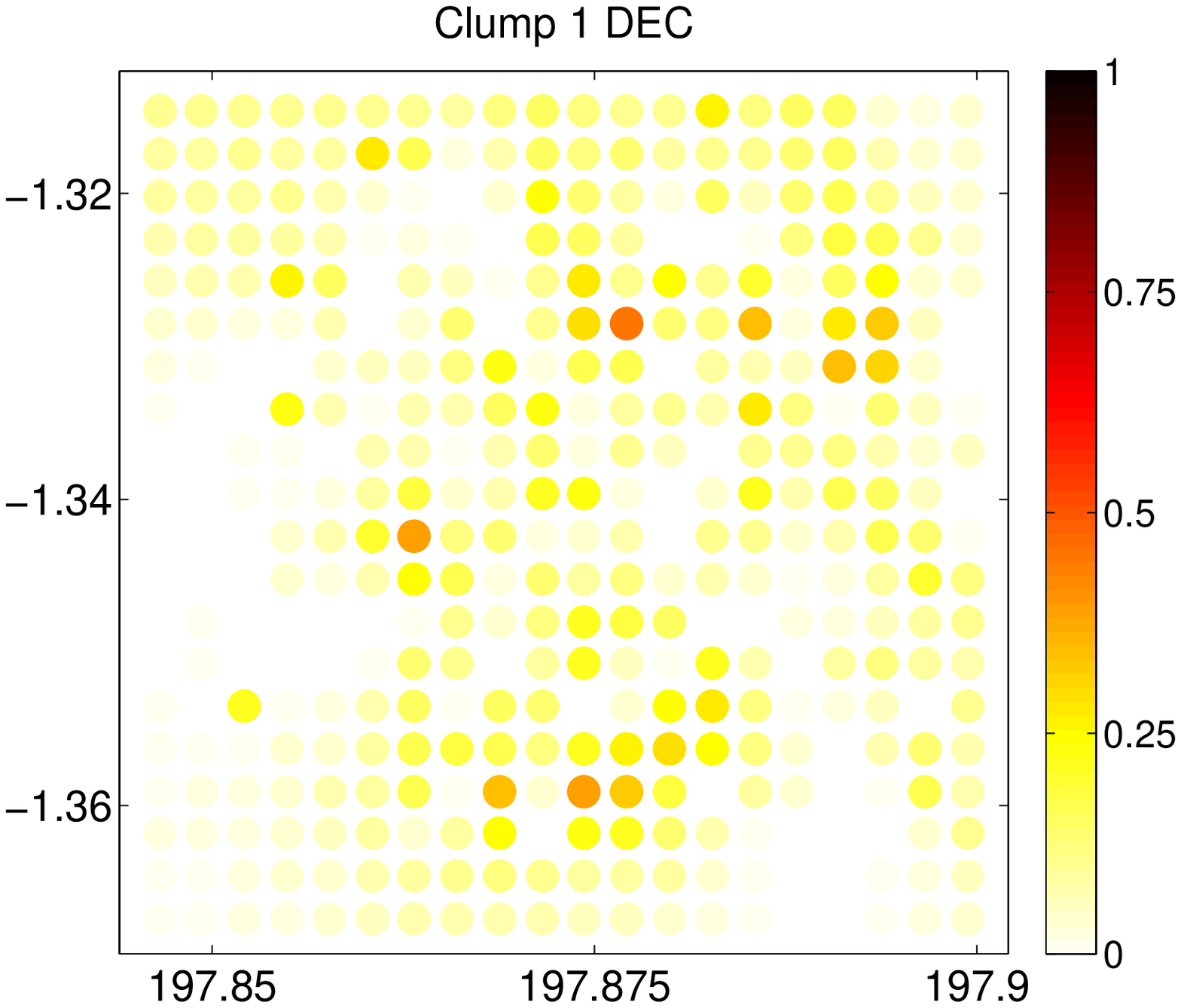,width=0.26\linewidth,clip=} & 
\epsfig{file=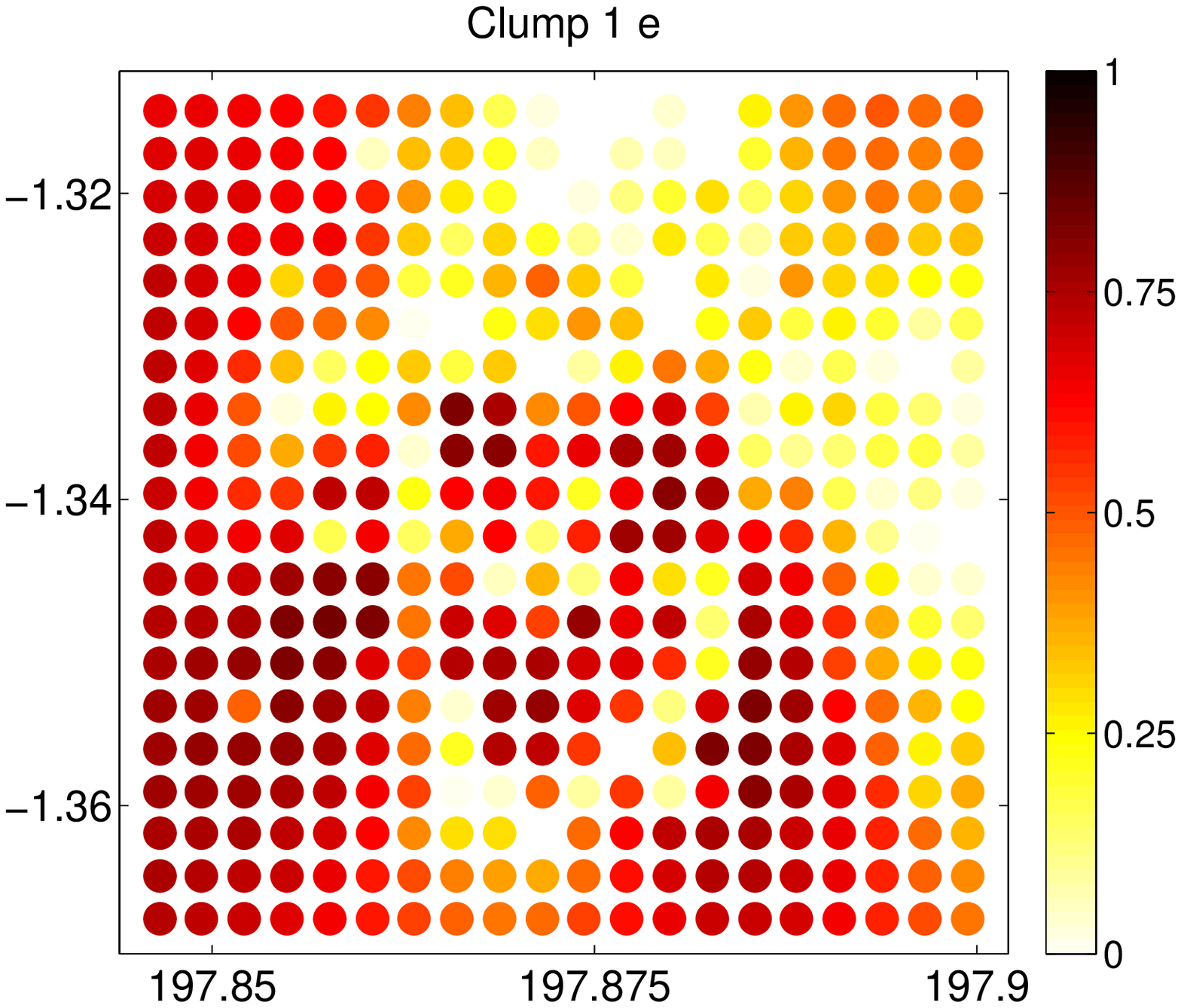,width=0.26\linewidth,clip=} \\
\epsfig{file=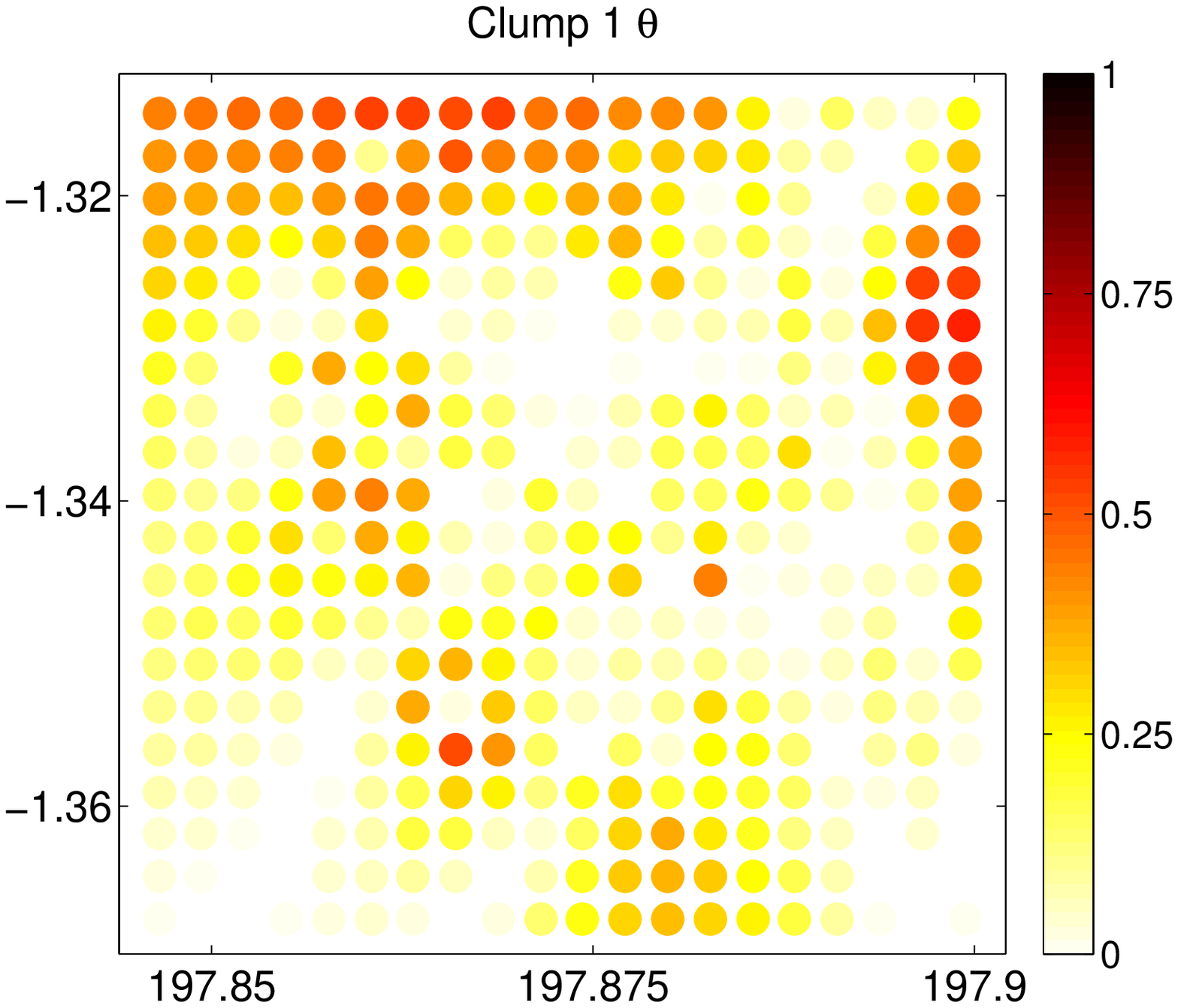,width=0.26\linewidth,clip=} &
\epsfig{file=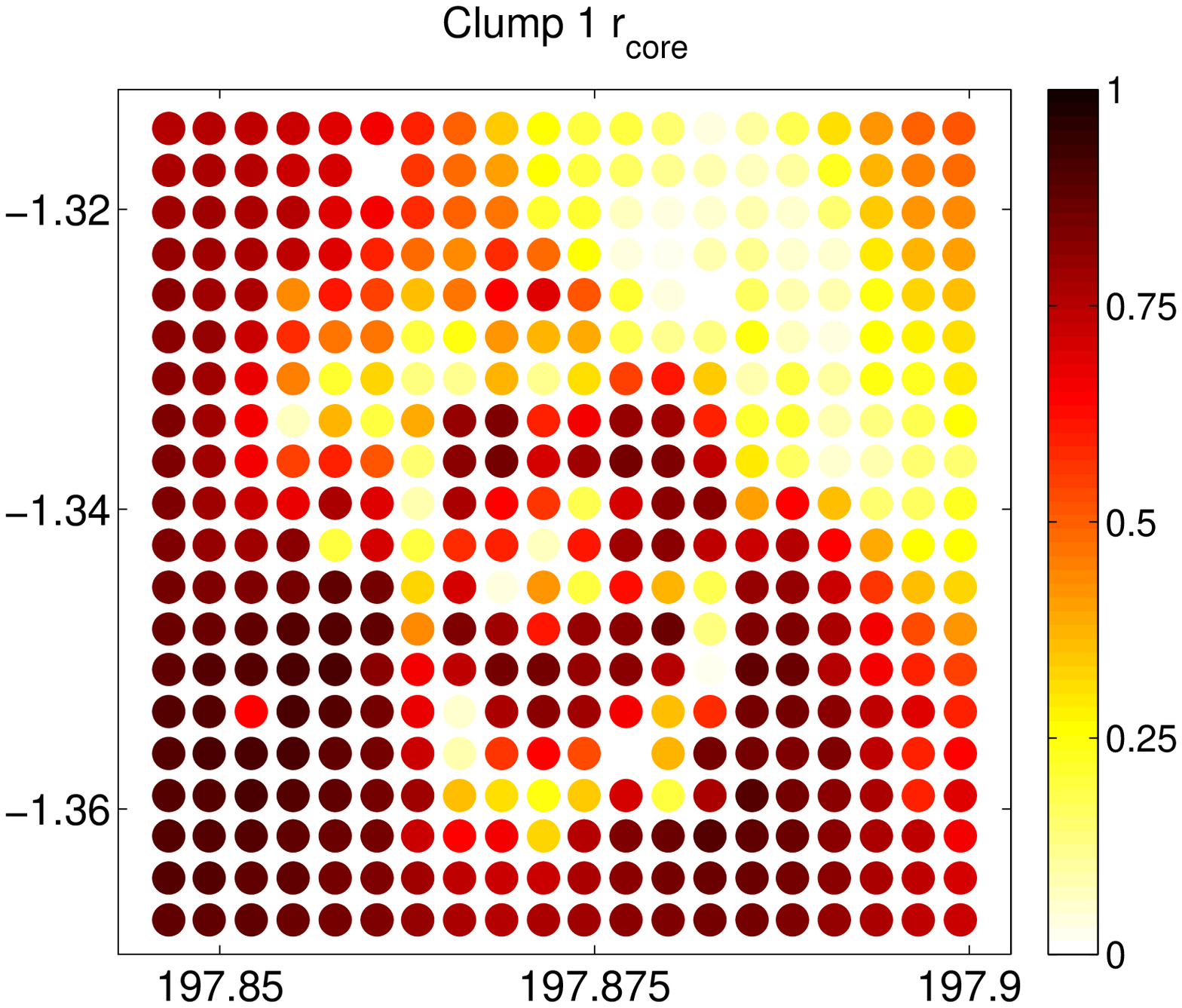,width=0.26\linewidth,clip=} & 
\epsfig{file=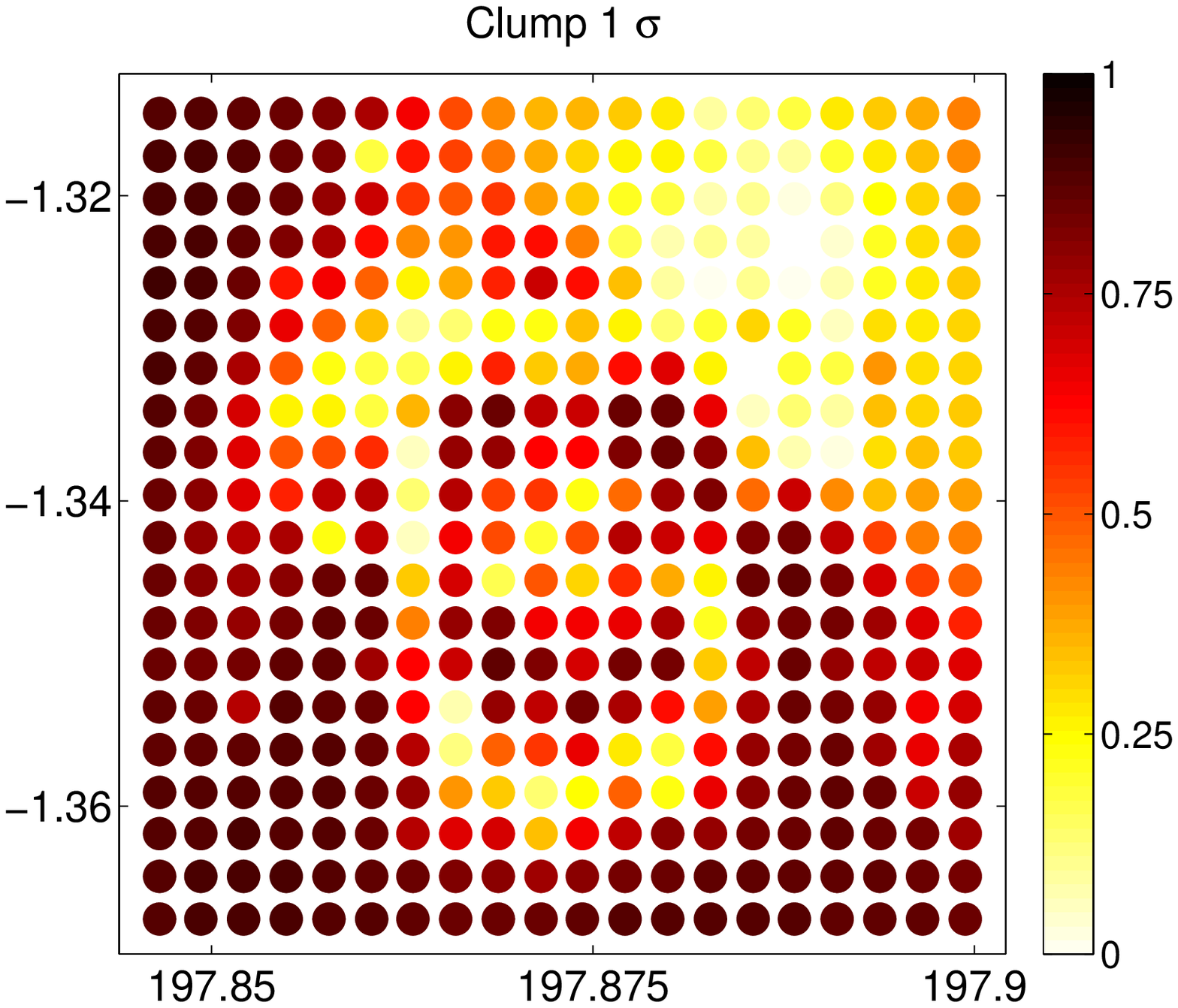,width=0.26\linewidth,clip=} \\
\epsfig{file=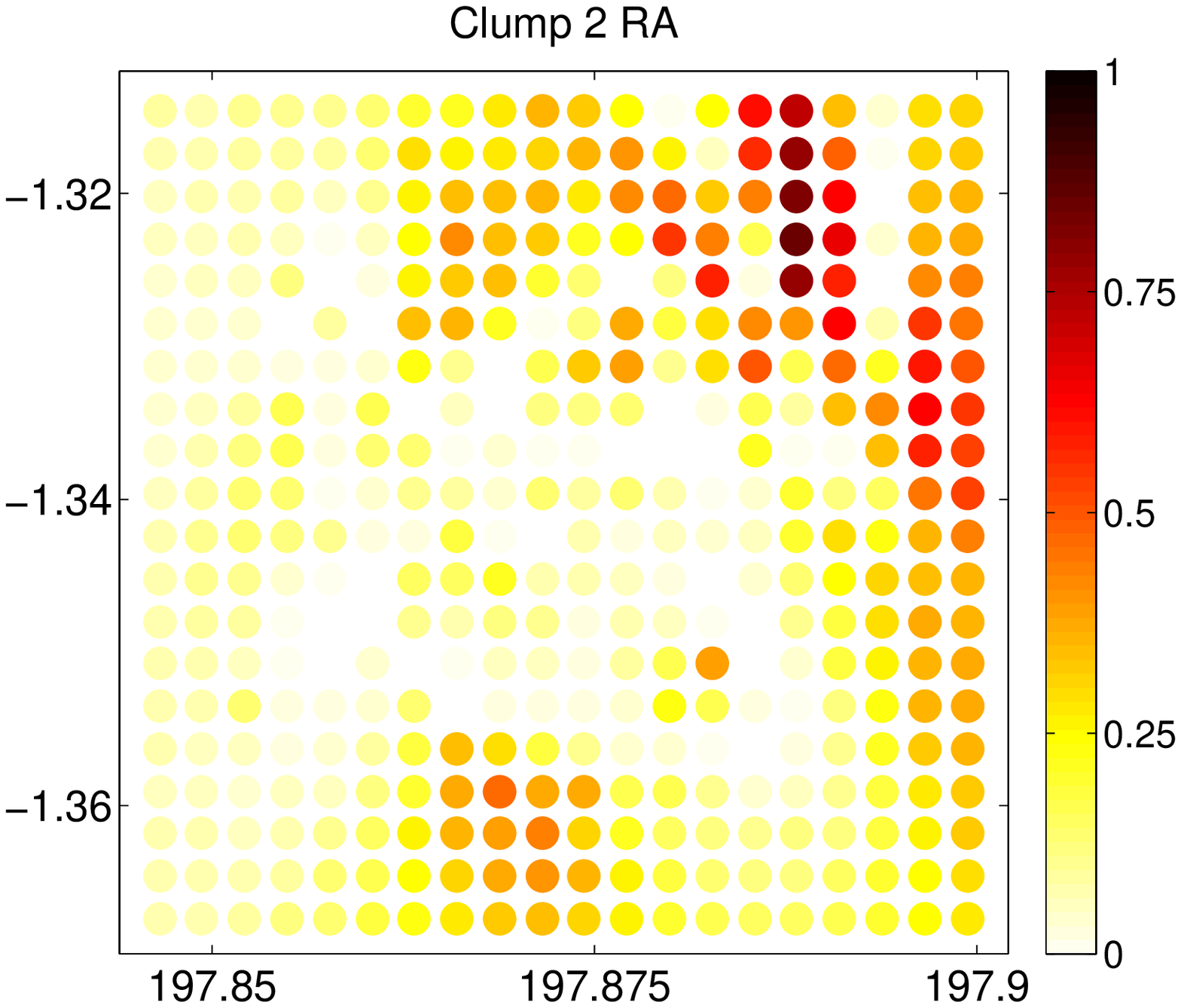,width=0.26\linewidth,clip=} & 
\epsfig{file=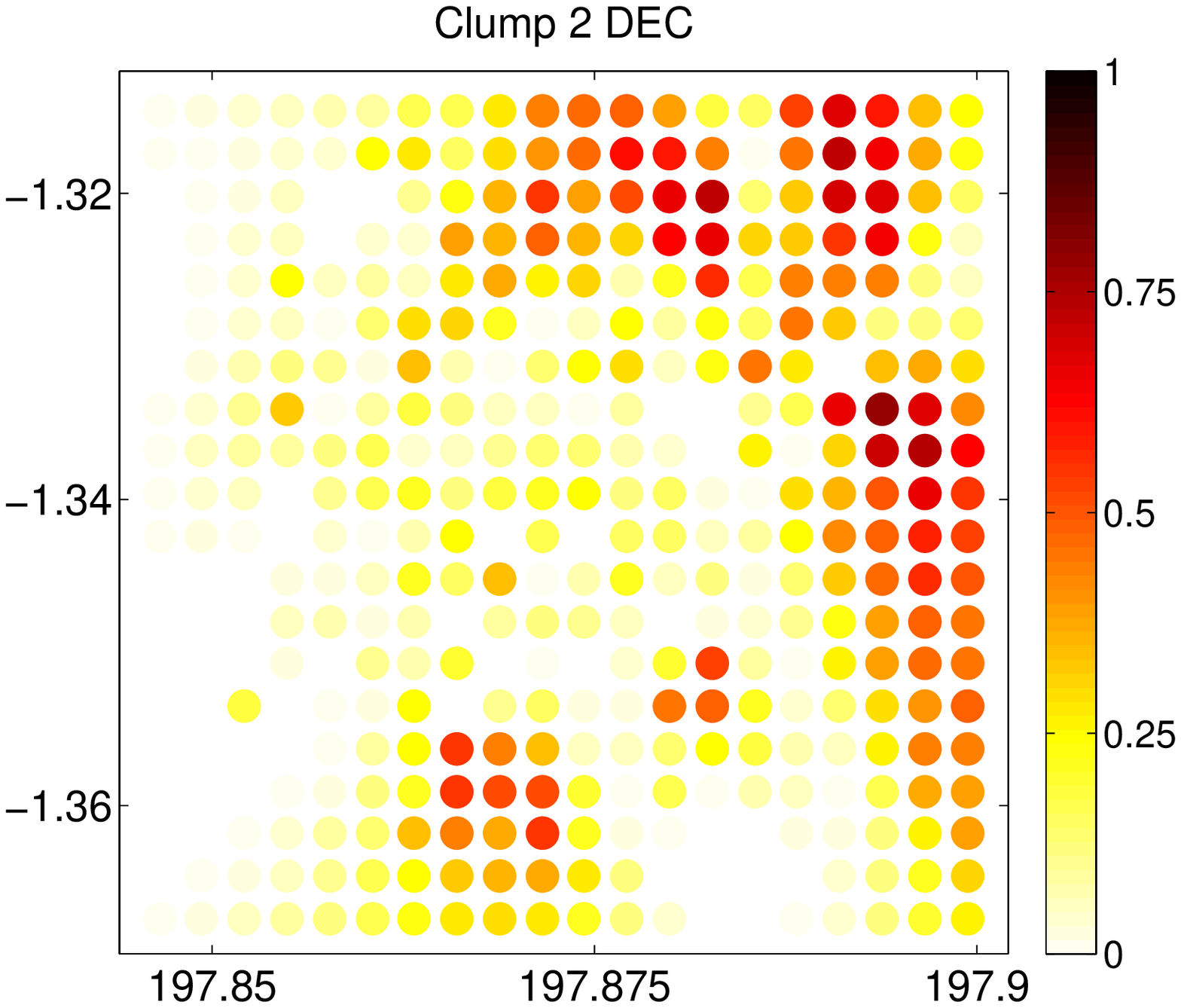,width=0.26\linewidth,clip=} & 
\epsfig{file=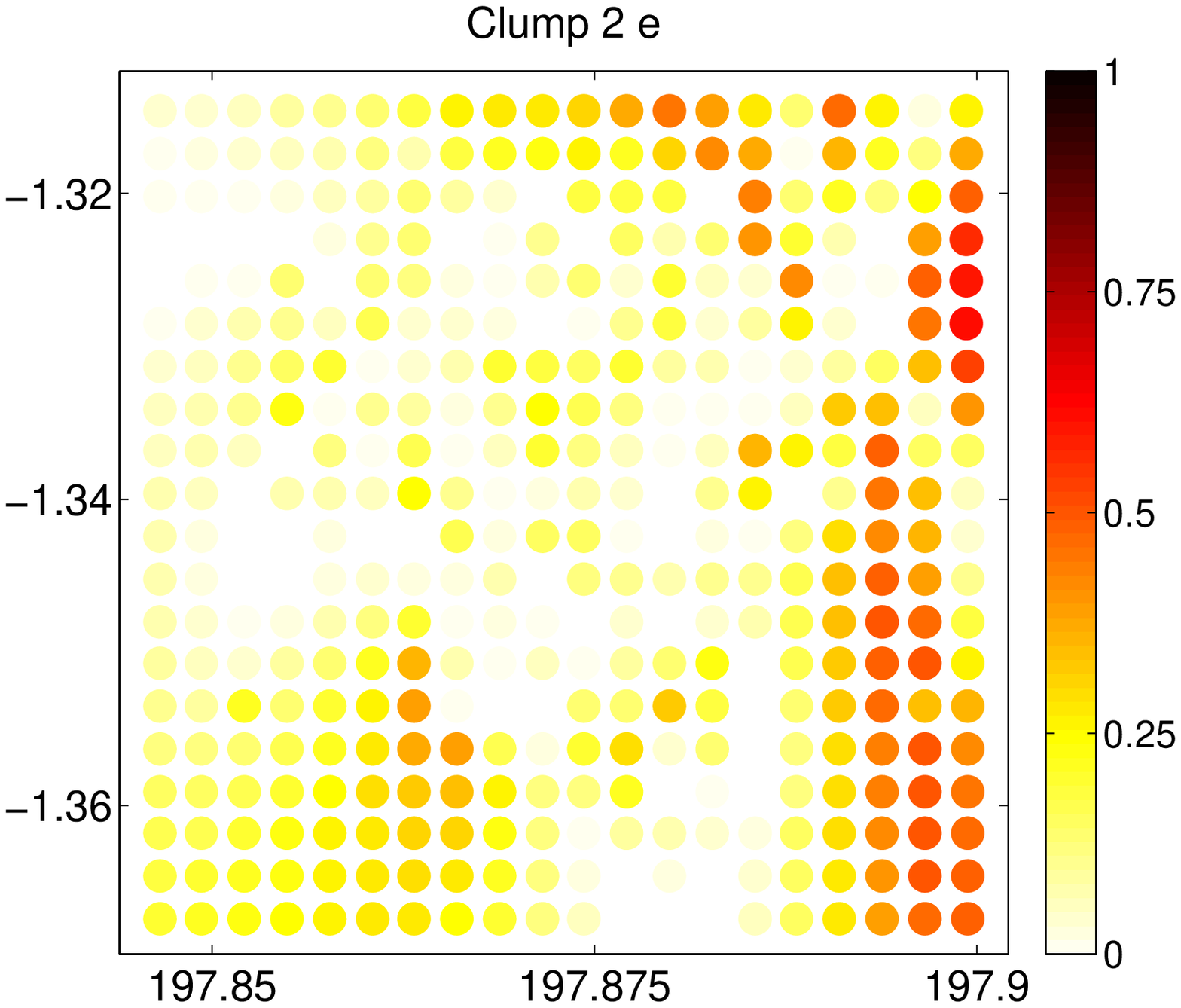,width=0.26\linewidth,clip=} \\
\epsfig{file=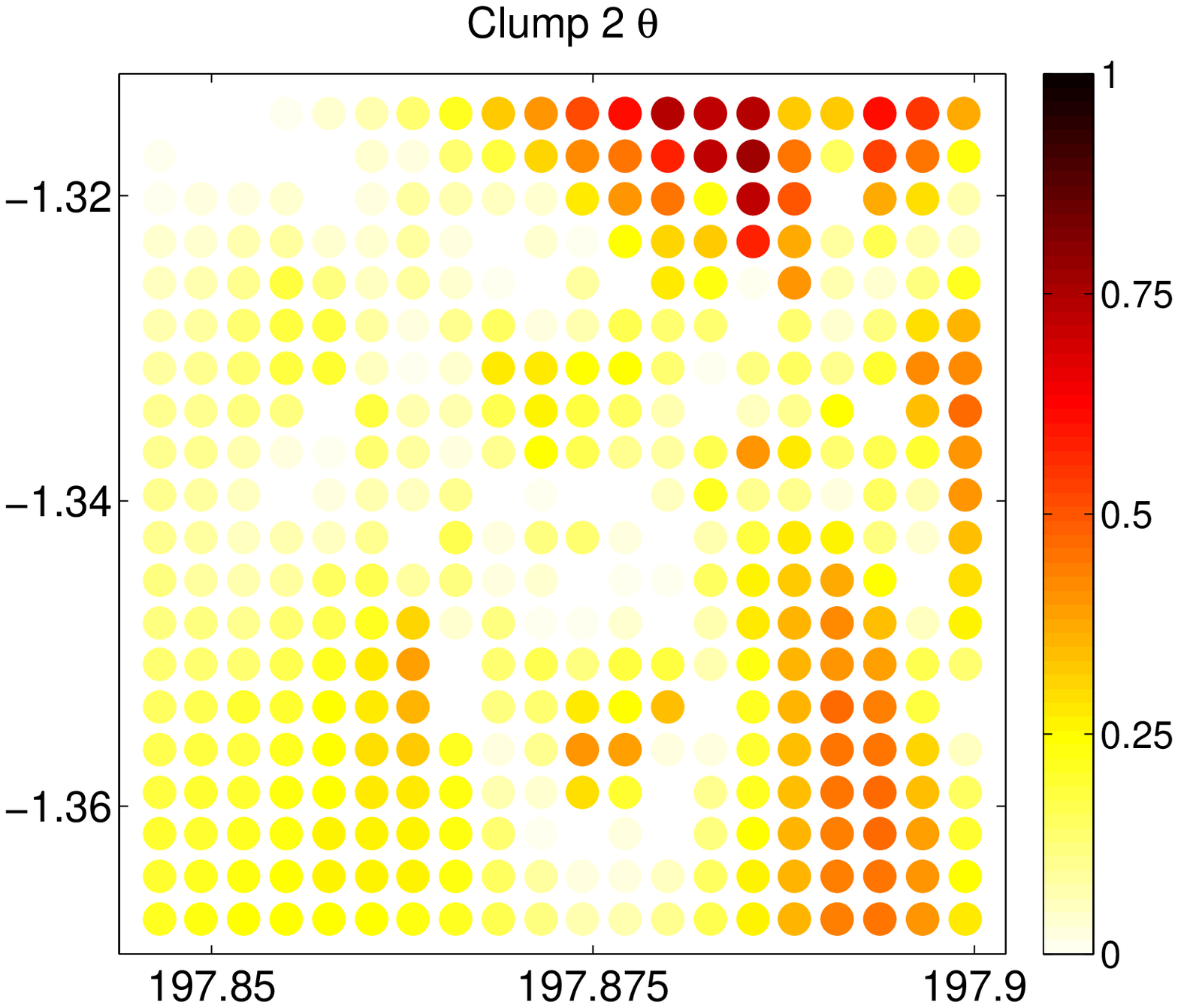,width=0.26\linewidth,clip=} & 
\epsfig{file=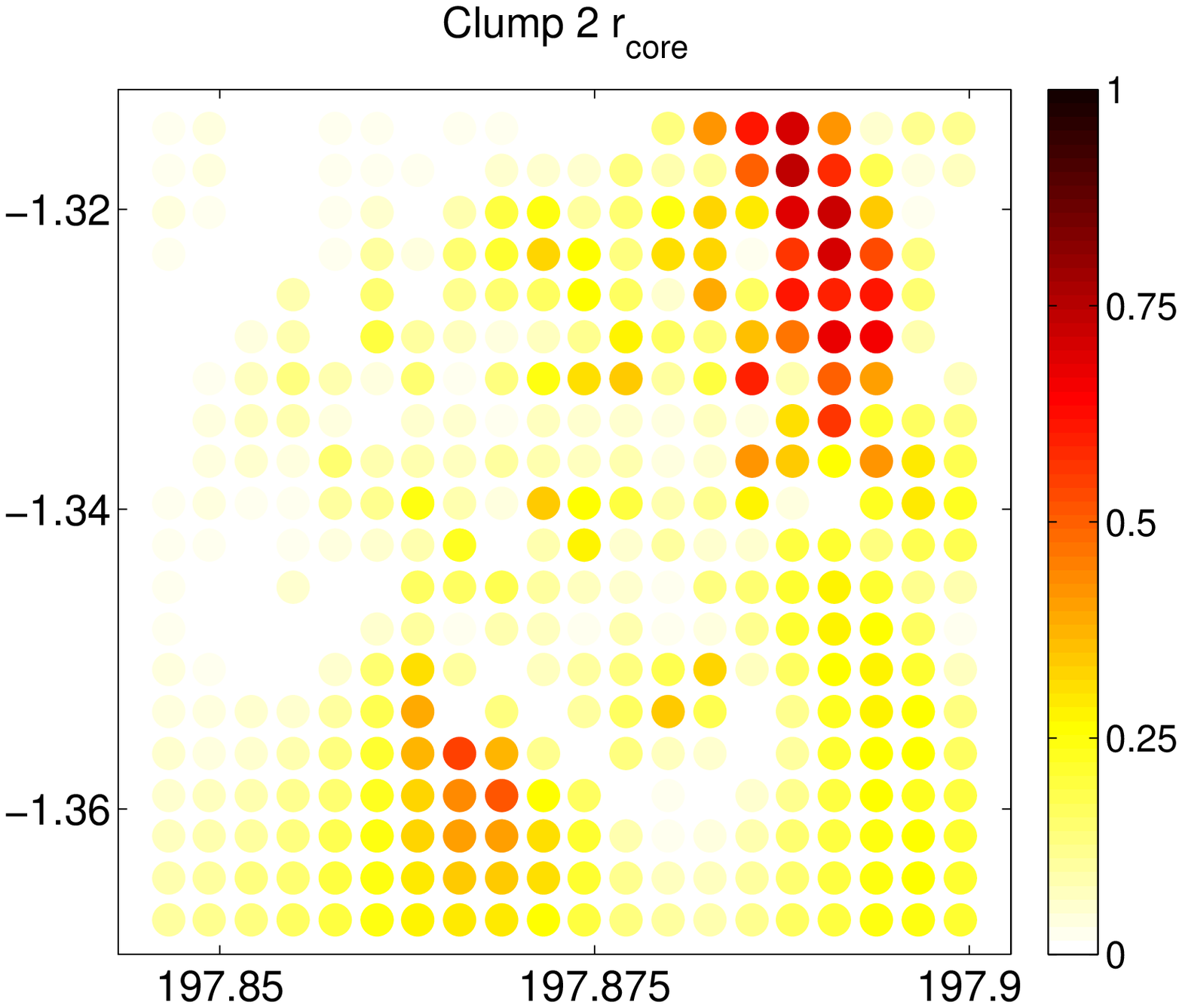,width=0.26\linewidth,clip=} & 
\epsfig{file=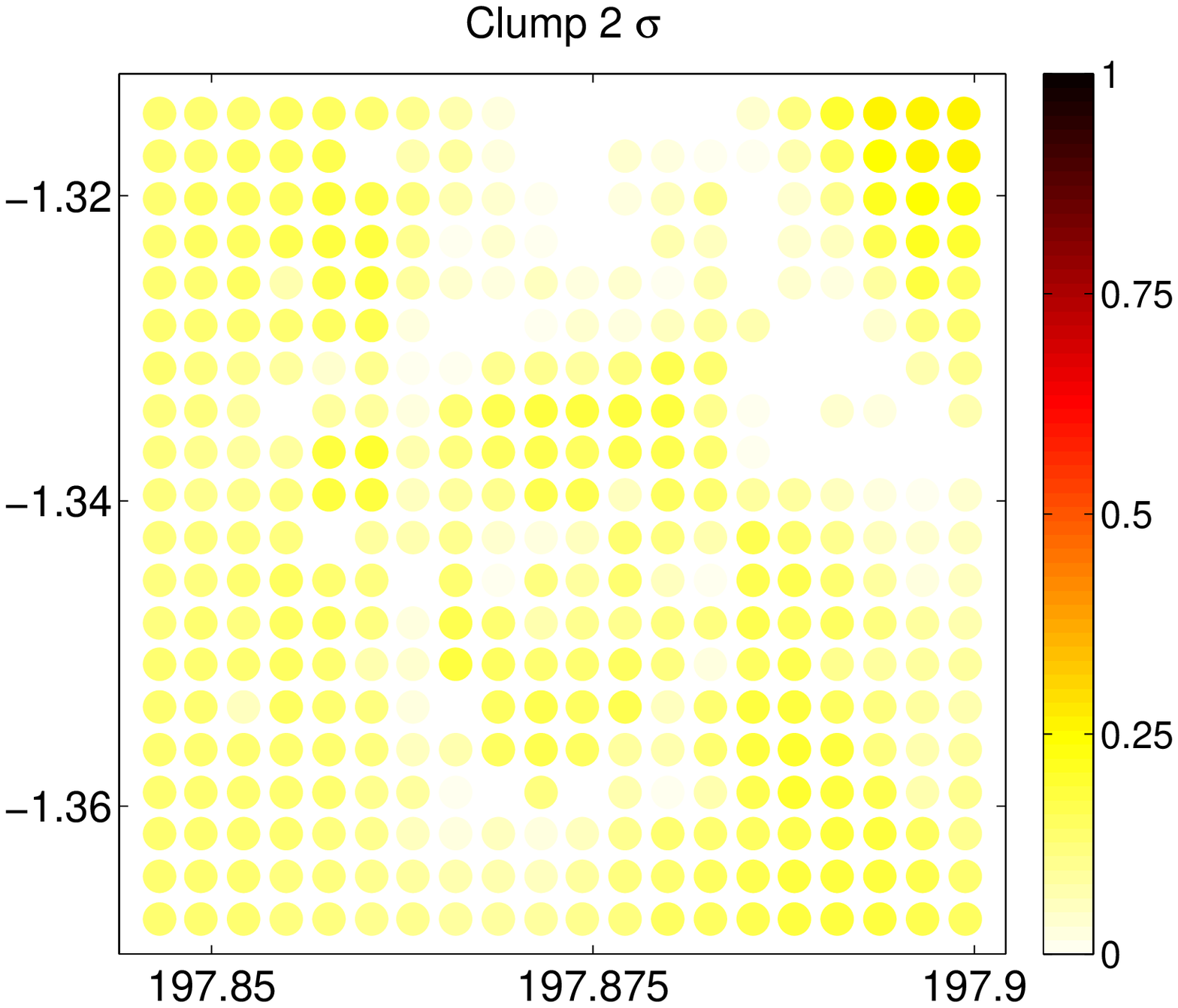,width=0.26\linewidth,clip=} \\
\epsfig{file=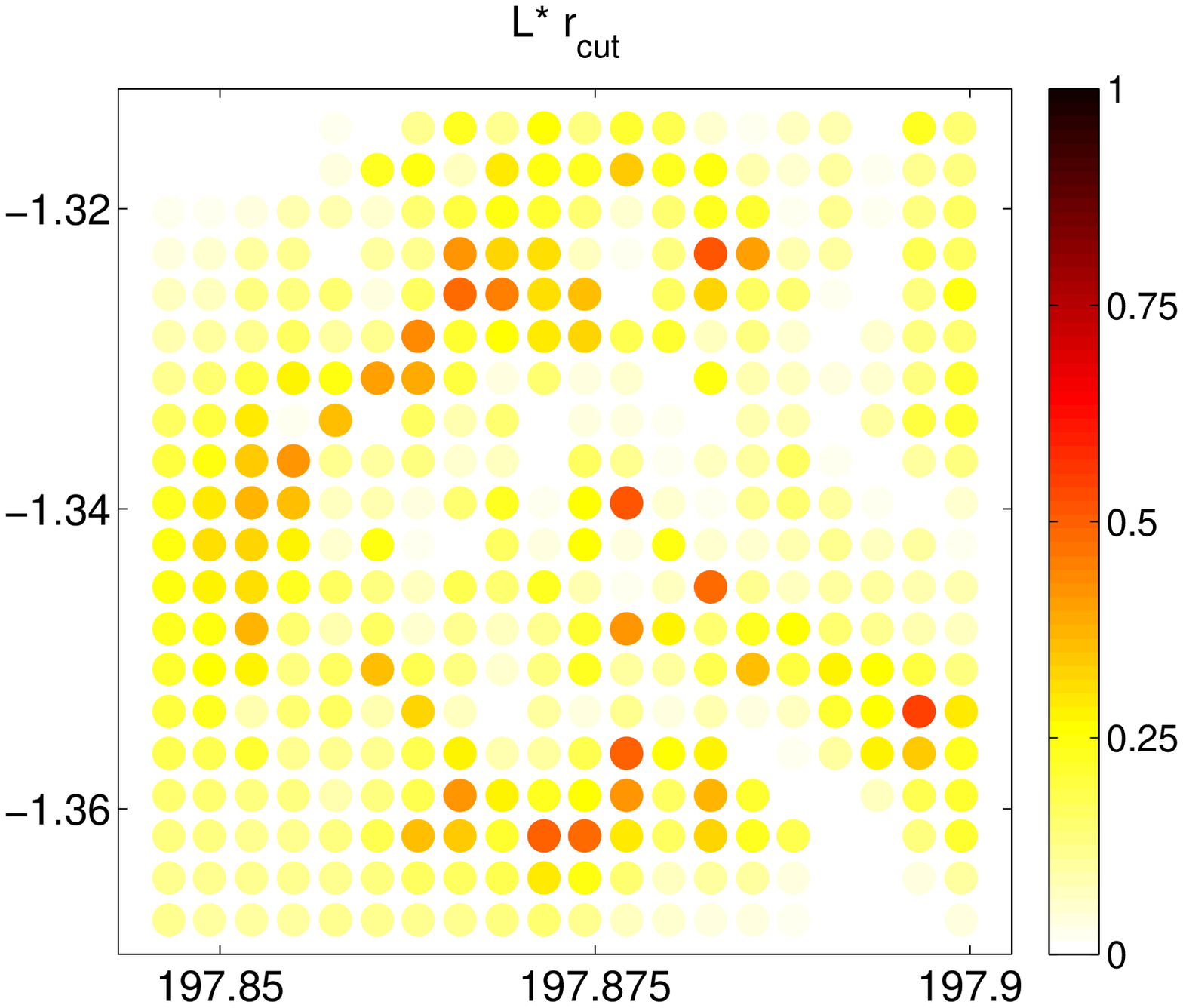,width=0.26\linewidth,clip=} & 
\epsfig{file=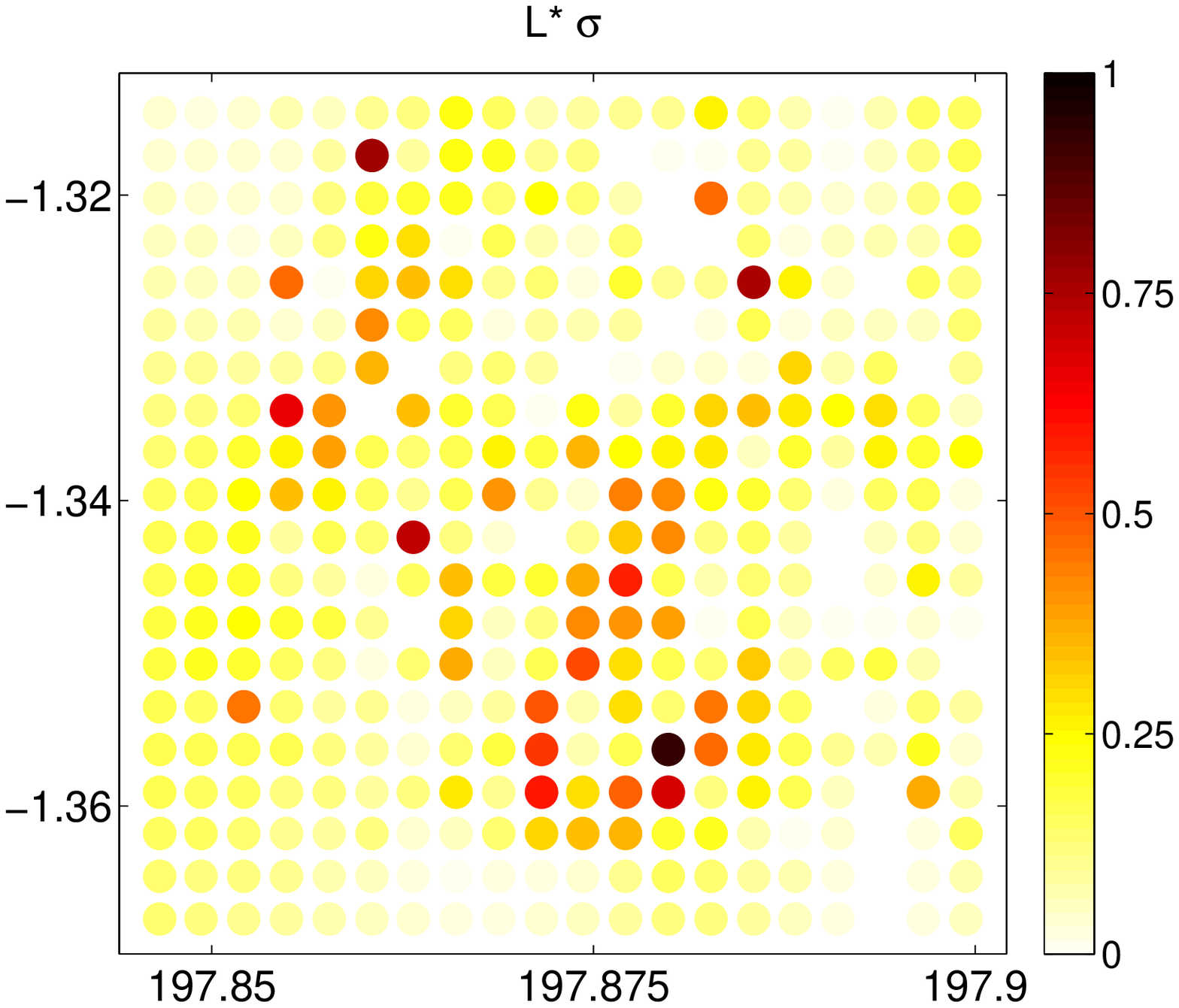,width=0.26\linewidth,clip=} &
\epsfig{file=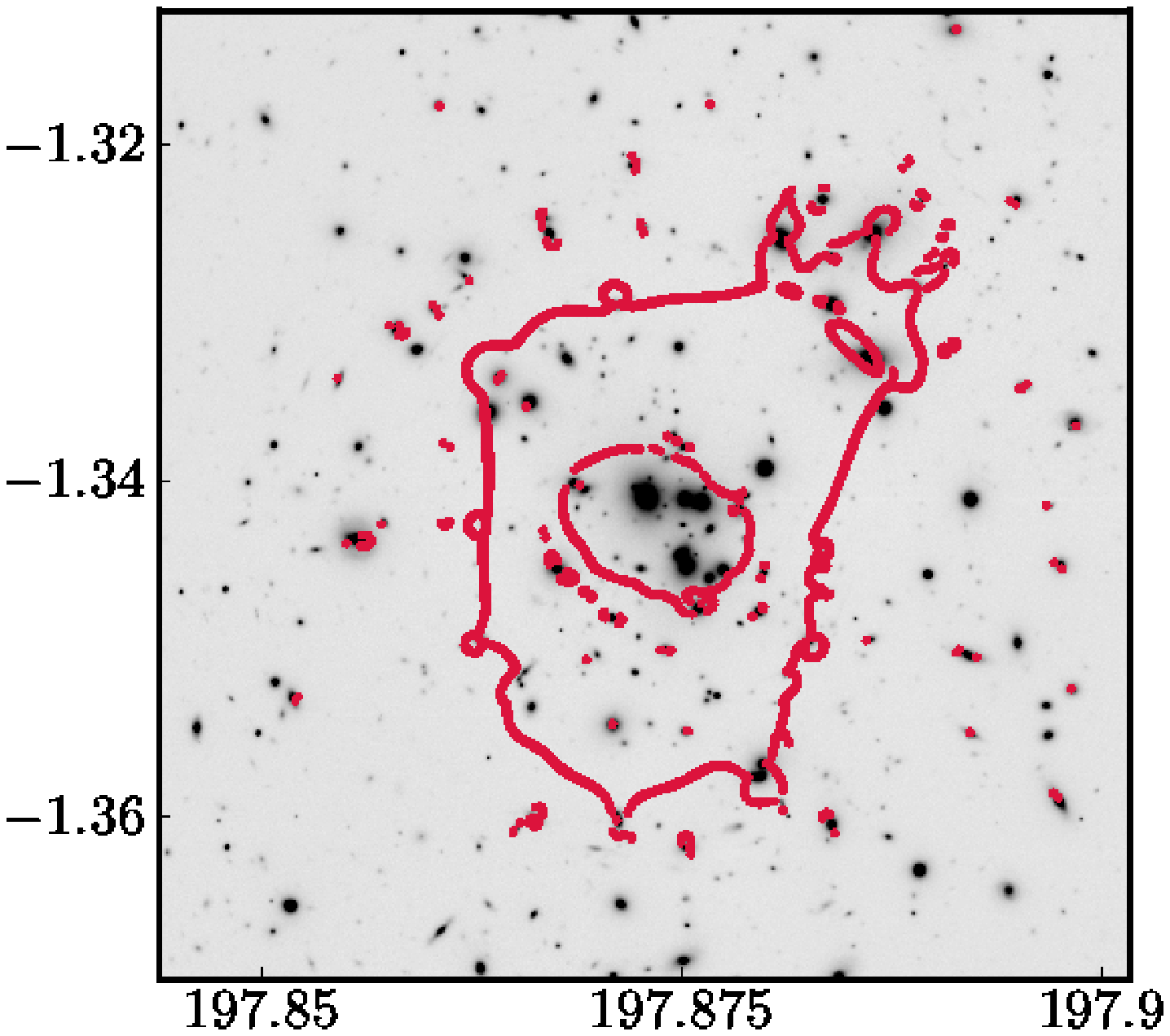,width=0.275\linewidth,clip=}
\end{tabular}

\end{figure*}

\begin{figure*}
\caption{Possible constraints on the model parameters describing the 
velocity dispersion (left) and core radius (right) of the main dark 
matter clump of A1689 using information on the magnification from 
a SN~Ia observed in the
multiply lensed background galaxy image 18.1 at $z$ = 1.82. The shaded
bars represent the parameter distribution from all realizations while
the unfilled bars give the parameter distribution considering only
realizations which reproduce the observed magnification within $\pm$
0.1 mag.}
\label{fig:cuts}
\centering
\begin{tabular}{cc}
\epsfig{file=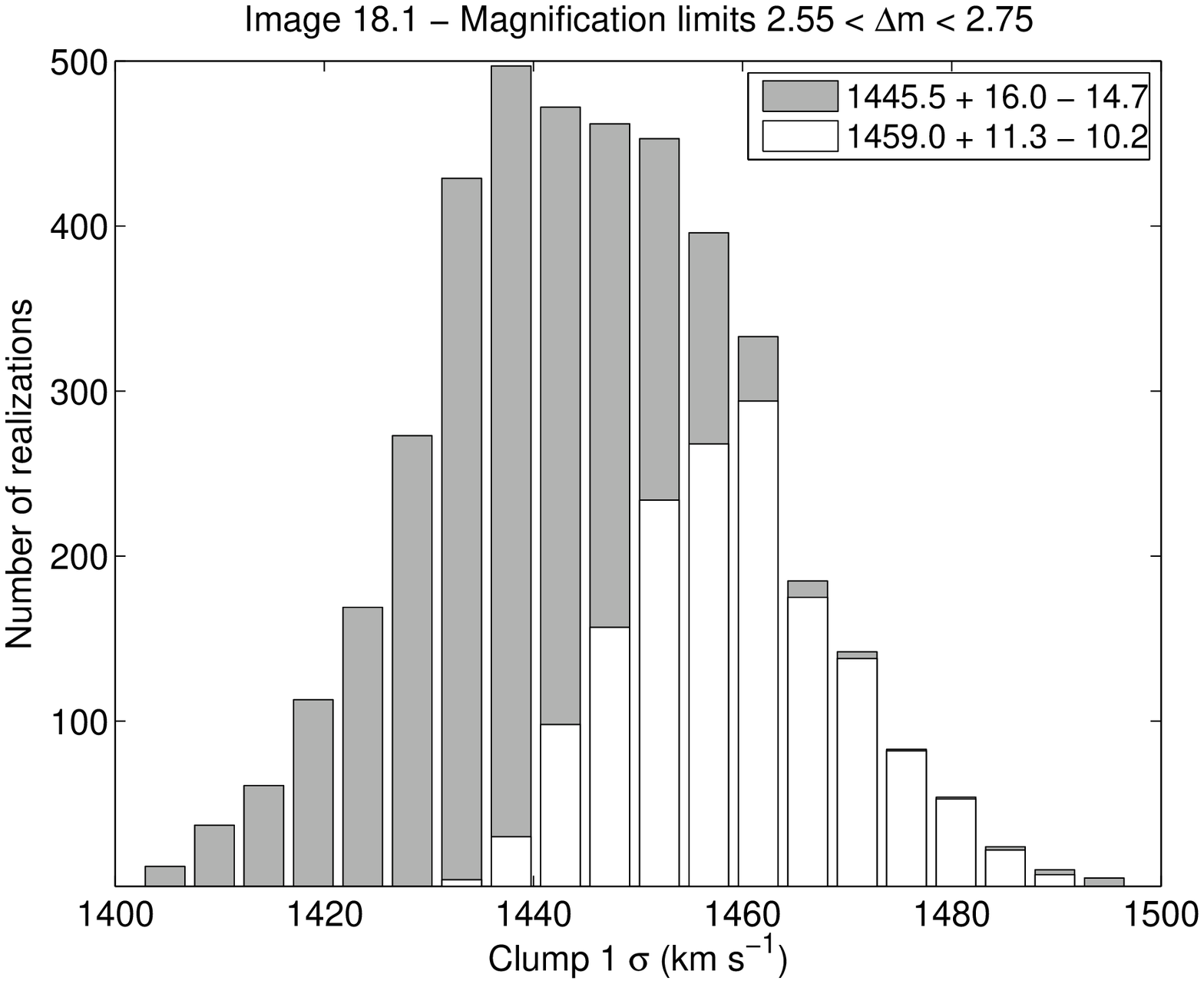,width=0.48\linewidth,clip=} & 
\epsfig{file=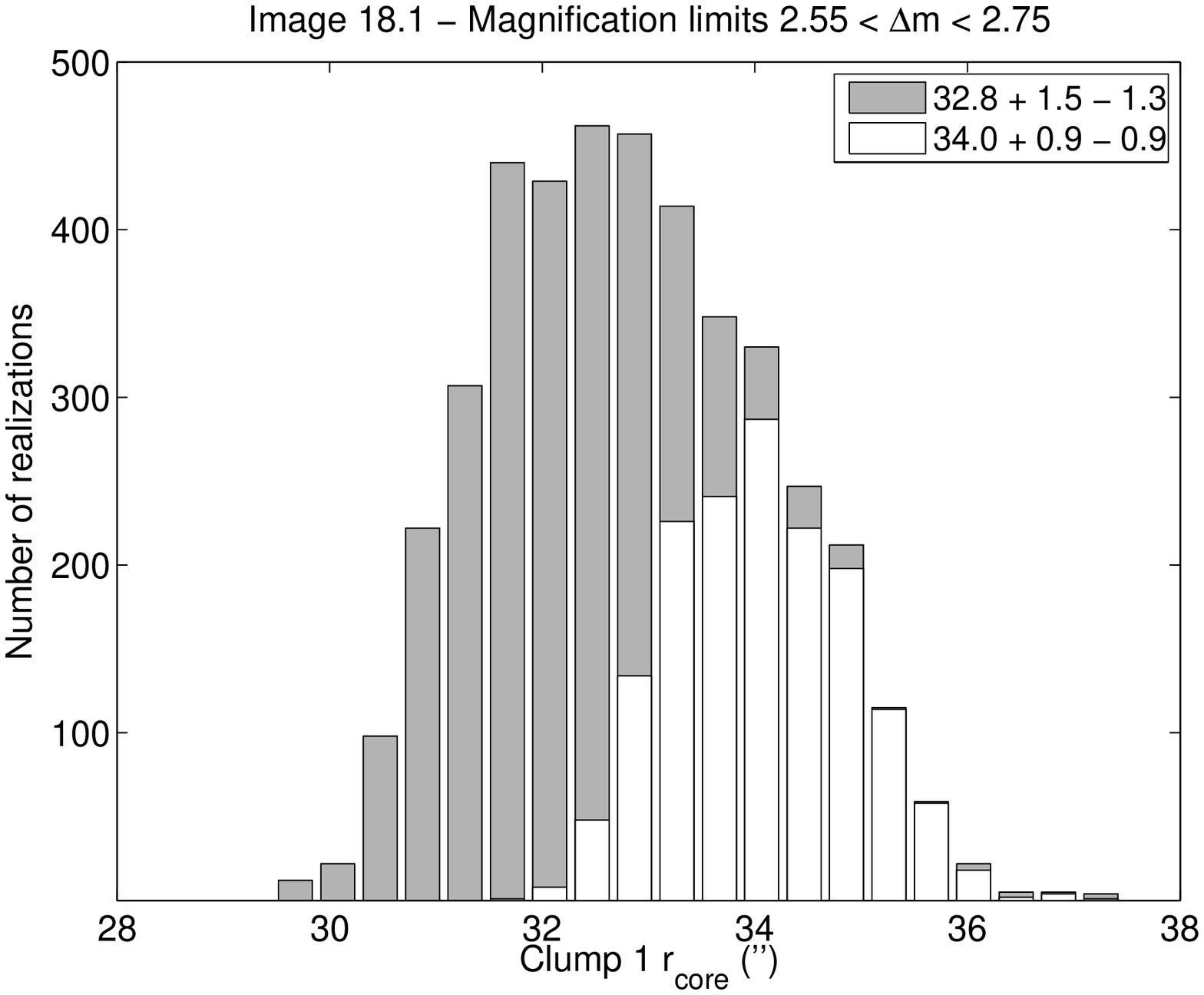,width=0.48\linewidth,clip=}  
\end{tabular}

\end{figure*}

Figure~\ref{fig:z2correls} shows the absolute value of the correlation
coefficients for the predicted magnification and the free input
parameters as a function of position for an image at redshift $z = 2$.
The parameters shown are those describing the two dark matter clumps (Clump 1 and Clump 2) and the cluster galaxy scaling relations ($L^*r_{\rm cut}$ and
$L^*\sigma$) in the model of A1689. A
value close to 1 implies a very strong correlation while a correlation
coefficient close to 0 implies no correlation. As can be seen in
Fig.~\ref{fig:z2correls}, the free parameters differ in the strength
of their correlation with the predicted magnification both with each
other and the position behind the cluster. The dependence of the
correlation strength on redshift is very weak. In general, of all
33 free optimized parameters, the ones which show the strongest
correlations and therefore the possibility of improving the model once
a SN~Ia is observed, are the ellipticity, $e$, core radius, $r_{\rm core}$, and
velocity dispersion, $\sigma$, of the main lensing potential (Clump
1). Other parameters, like the corresponding parameters for the
cluster galaxies and the position of the main lensing potentials show
a weaker correlation and can probably not be further constrained using
SN~Ia observations in the case of A1689.

Figure~\ref{fig:cuts} shows an example of how the model parameters
could be constrained if the magnification of an image is known to a precision 
of 0.1 mag. In this example we asssume a SN~Ia observed in the 
multiply-lensed
image 18.1 with spectroscopic redshift $z$ = 1.82 under the most
favorable conditions. Since this image does not have a counter image with a 
time delay below five years, it is not included in Table~\ref{tab:spec_delays}. 
At the position of this image, there are strong
correlations between the model parameters and the predicted
magnification of this image. The correlation coefficient 
for the velocity dispersion and  core radius of the main clump (Clump 1)
are 0.84 and 0.87, respectively. 
As can
be seen in Figure~\ref{fig:cuts}, the possible constraints on $\sigma$
for the main lensing potential improve
from $\sim 1\%$ to $\sim 0.7\%$, while the corresponding numbers for
$r_{\rm core}$ are $\sim 4.5\%$ and $\sim 2.6\%$. Thus,
adding only one constraint from the observed magnification of a SN~Ia,
the constraints on the model parameters for this already very well
constrained cluster model of A1689 (114 multiple images and 24
spectroscopic redshifts), improve by almost a factor of 2. It should
however be noted, that the possible constraints from this image should
be seen as a best case scenario, as the correlations for other systems
are typically weaker, with median absolute correlation coefficient
values in a FOV extending $\pm$ 100 arcseconds around the center of the cluster
at $z = 2$ of 0.67 and 0.71 for the velocity dispersion of the main
clump (Clump 1) and its core radius,
respectively. Since we are expecting $\sim 5$ SNe~Ia exploding in
all background galaxies per year, we would be able to obtain a
combined constraint from such observations of the magnification at
several positions behind the cluster.

\begin{figure*}
\caption{Correlations between the predicted magnification of an 
image as a function of image position for a source at redshift $z = 1$ and the different input parameters which are
optimized in the model for A2204 (compare
Table~\ref{tab:A2204par} and Fig.~\ref{fig:z2correls}) in a field of view $\pm$ 100 arcseconds
around the cluster center. Correlations are given as the absolute
values of the correlation coefficients. The last panel shows an ACS
image of the cluster overlaid with the critical lines for $z = 1$.}
\label{fig:2204constr}
\centering
\begin{tabular}{ccc}
\epsfig{file=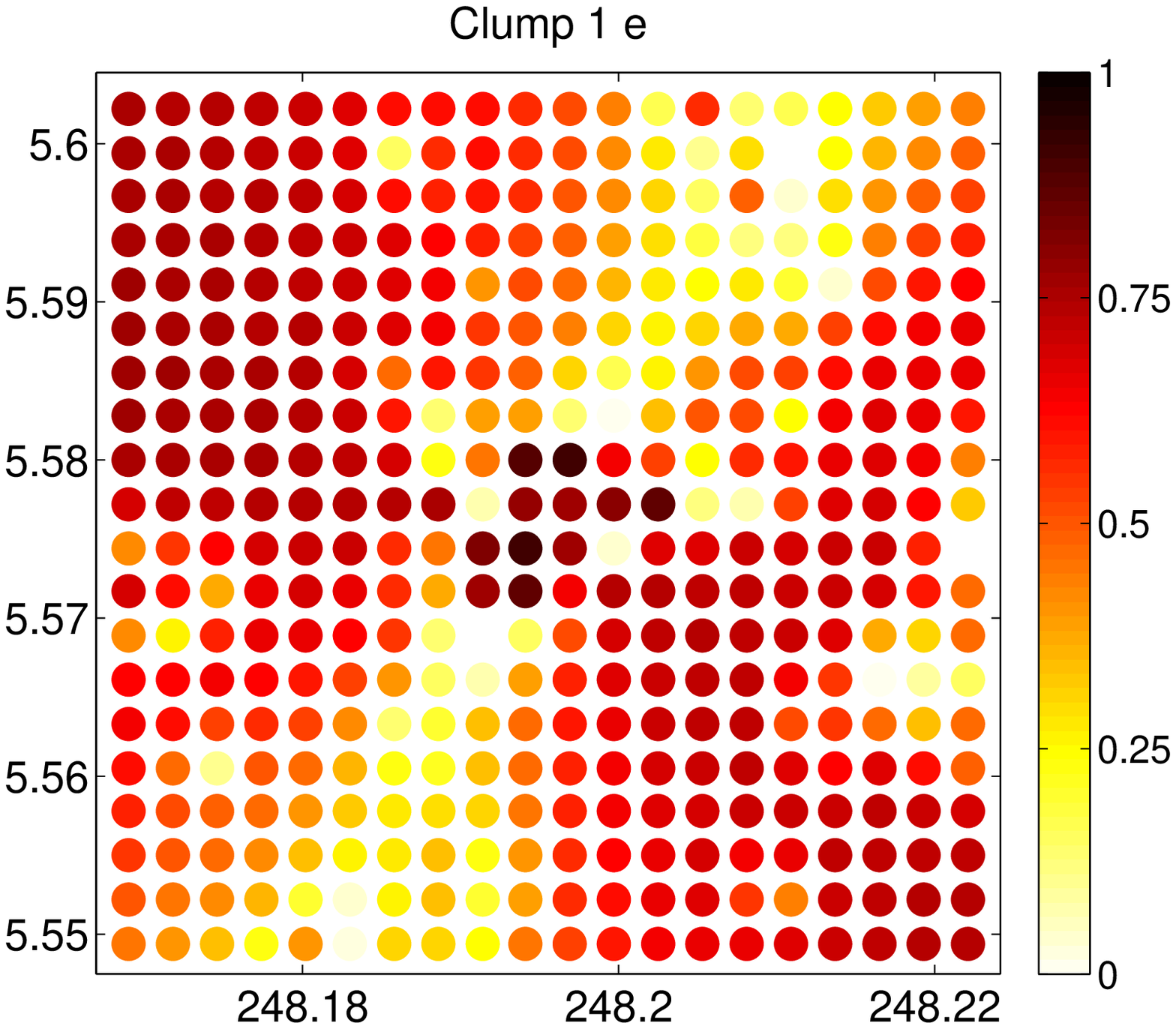,width=0.28\linewidth,clip=} & 
\epsfig{file=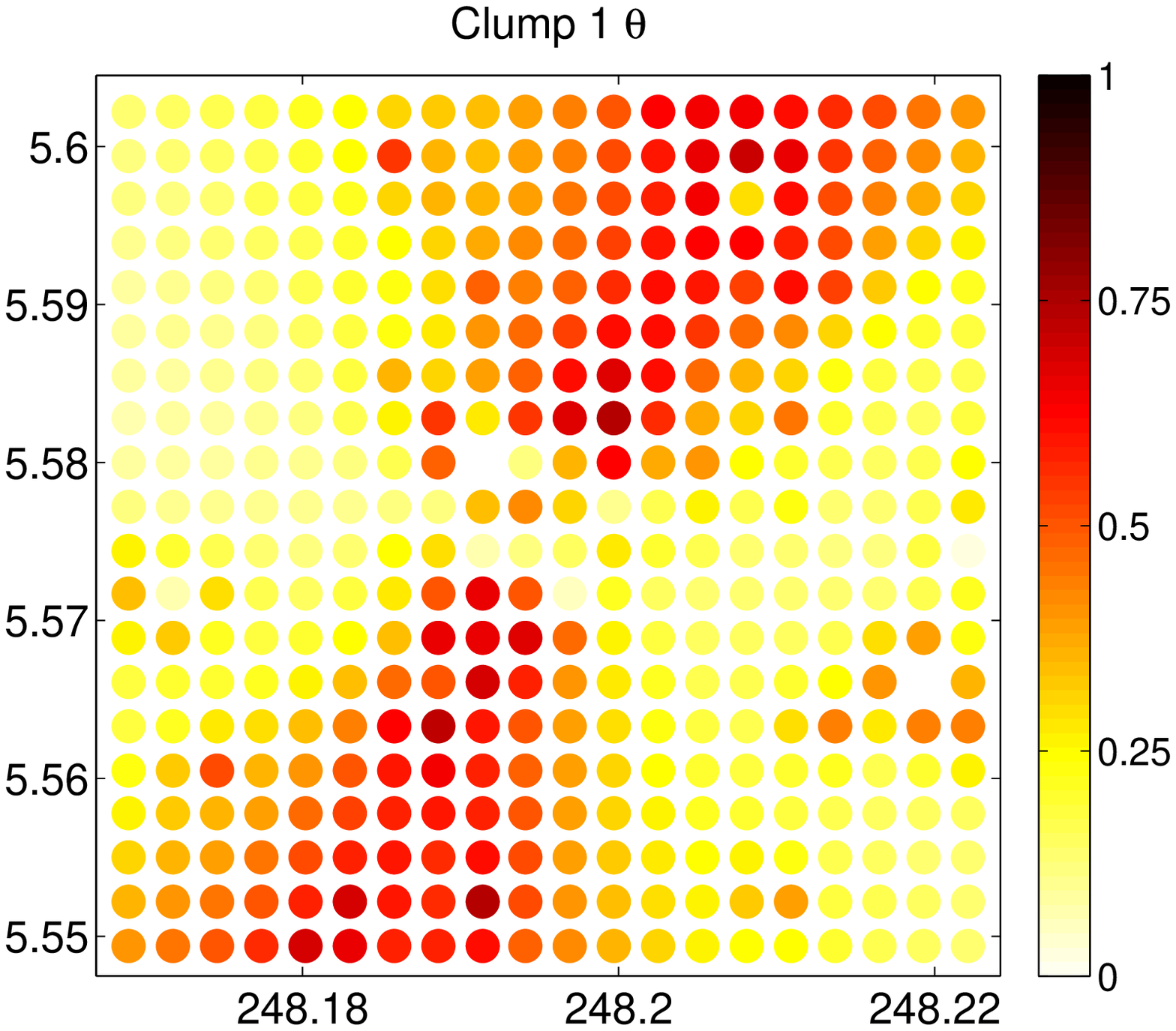,width=0.28\linewidth,clip=} & 
\epsfig{file=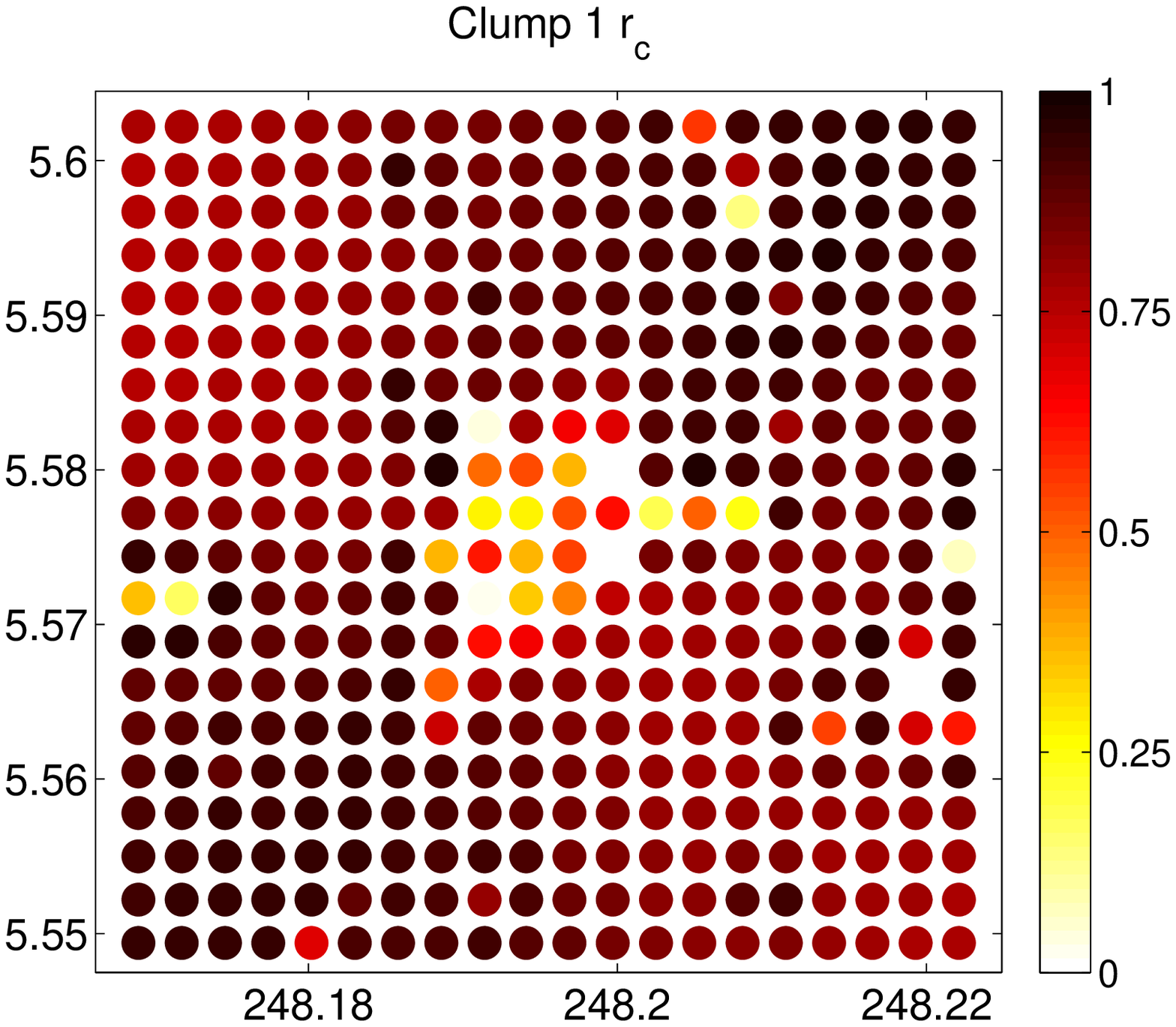,width=0.28\linewidth,clip=} \\
\epsfig{file=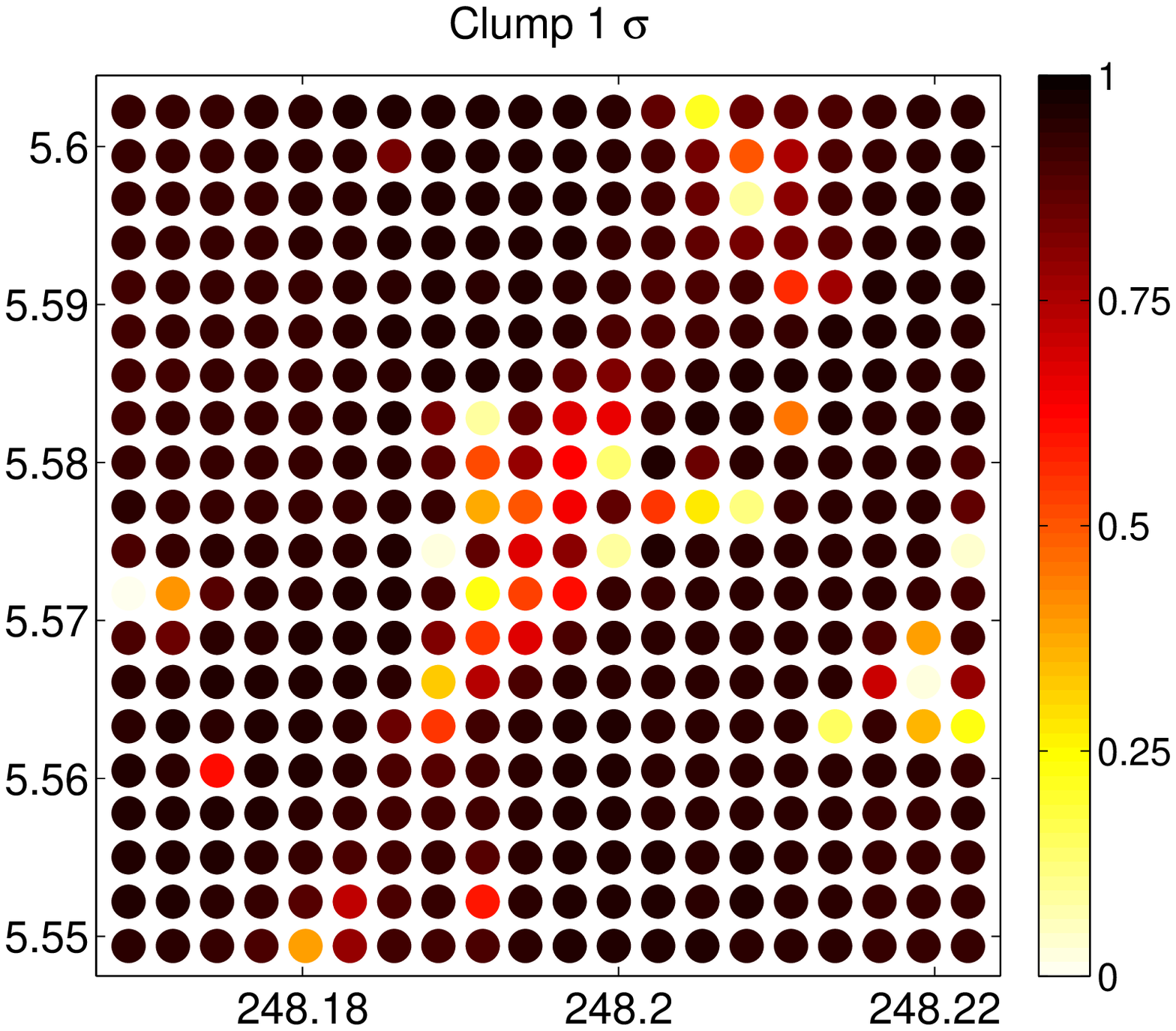,width=0.28\linewidth,clip=} &
\epsfig{file=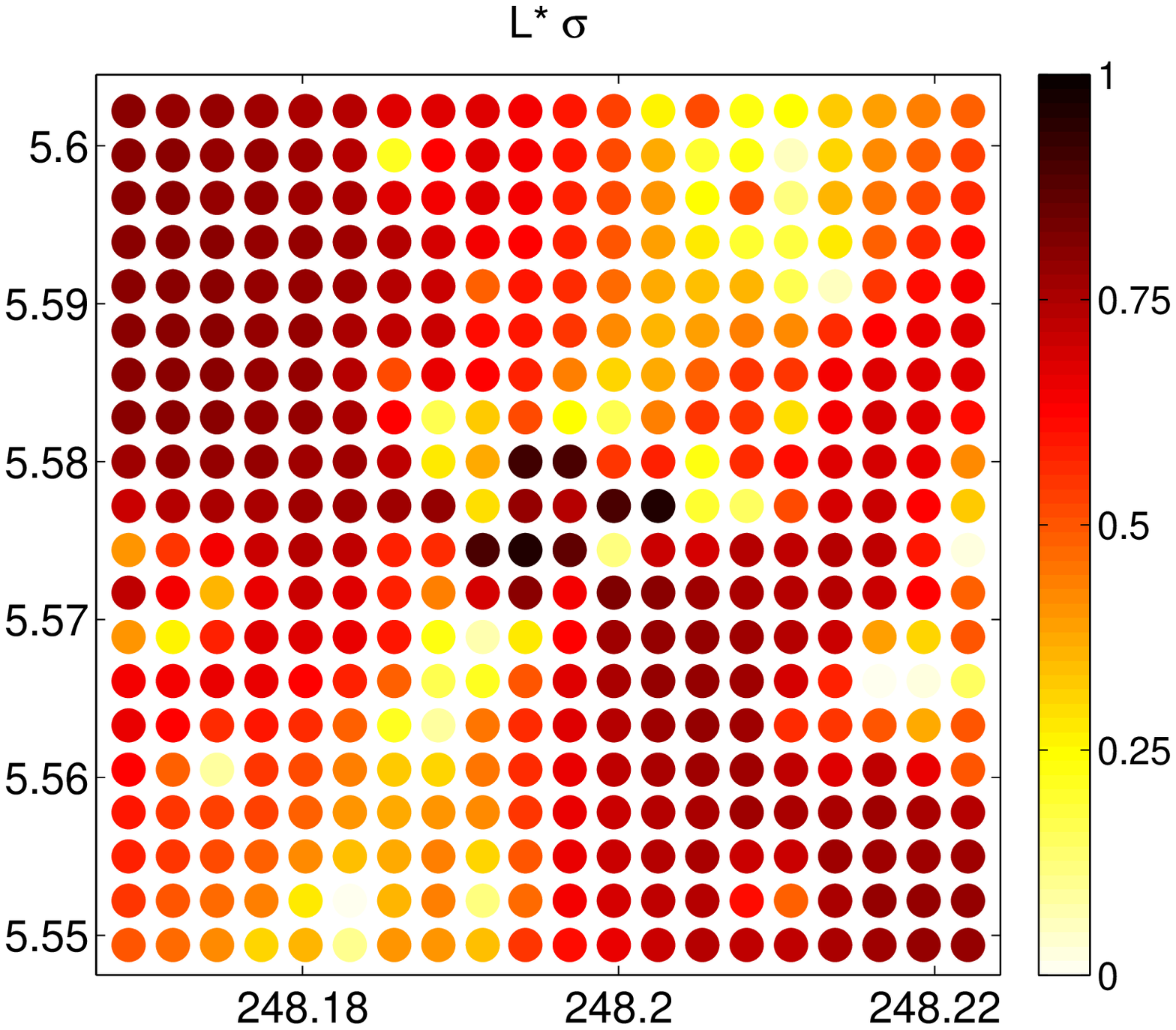,width=0.28\linewidth,clip=} &
\epsfig{file=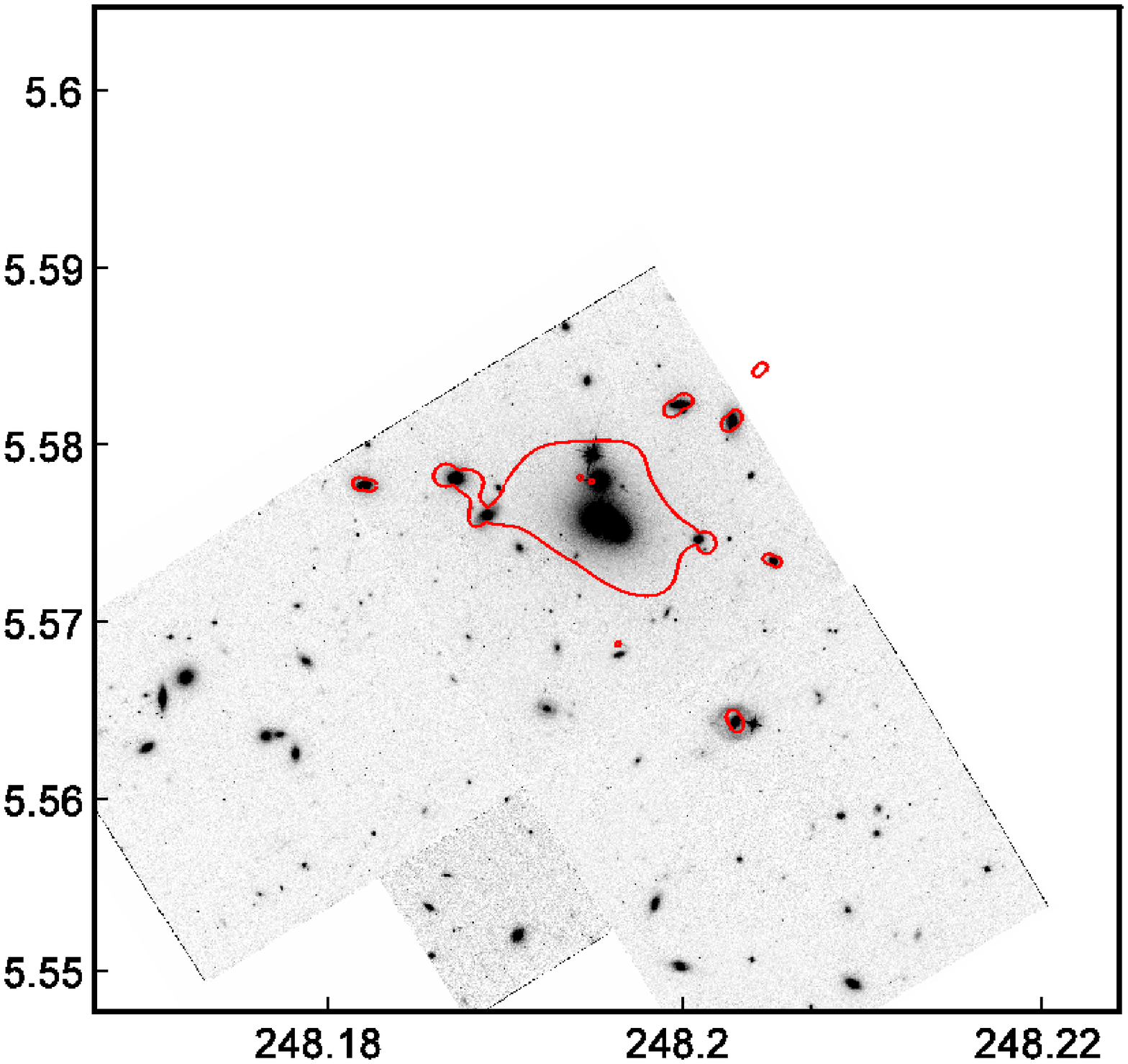,width=0.27\linewidth,clip=}
\end{tabular}

\end{figure*}

This technique might prove to be even more interesting for
testing the mass modeling of other less constrained galaxy
clusters. To evaluate the power of the method, we investigate
cluster A2204 with only one known
multiply lensed image system, making it much
less constrained than A1689, which we have been focusing on so far. Again,
we investigate the correlations of the optimized free parameters with
the resulting magnifications. The correlation strengths for a grid of
image positions $\pm$ 100 arcseconds around the cluster center at
redshift $z = 1$ is shown in Fig.~\ref{fig:2204constr}. Similarly to
the case of A1689, the strongest correlations are found for $\sigma$ and 
$r_{\rm core}$ of the main
potential with the median absolute correlation coefficient values in the FOV
shown in Fig.~\ref{fig:2204constr} at redshift $z = 1$
being 0.96 and 0.87, respectively. It is notable that these correlations
are rather weak in the central regions of the cluster potential, which
is constrained through the location of the known multiple images.

\begin{figure*}
\caption{Possible constraints on the model parameters describing the velocity dispersion (left) and core radius (right) of the main dark matter clump for 
A2204 using information on the magnification from a SN~Ia observed in
the image. The shaded bars represent the parameter distribution from
all realizations while the unfilled bars give the parameter
distribution considering only realizations which reproduce the
observed magnification within $\pm$ 0.1 mag.}
\label{fig:2204cuts}
\centering
\begin{tabular}{cc}
\epsfig{file=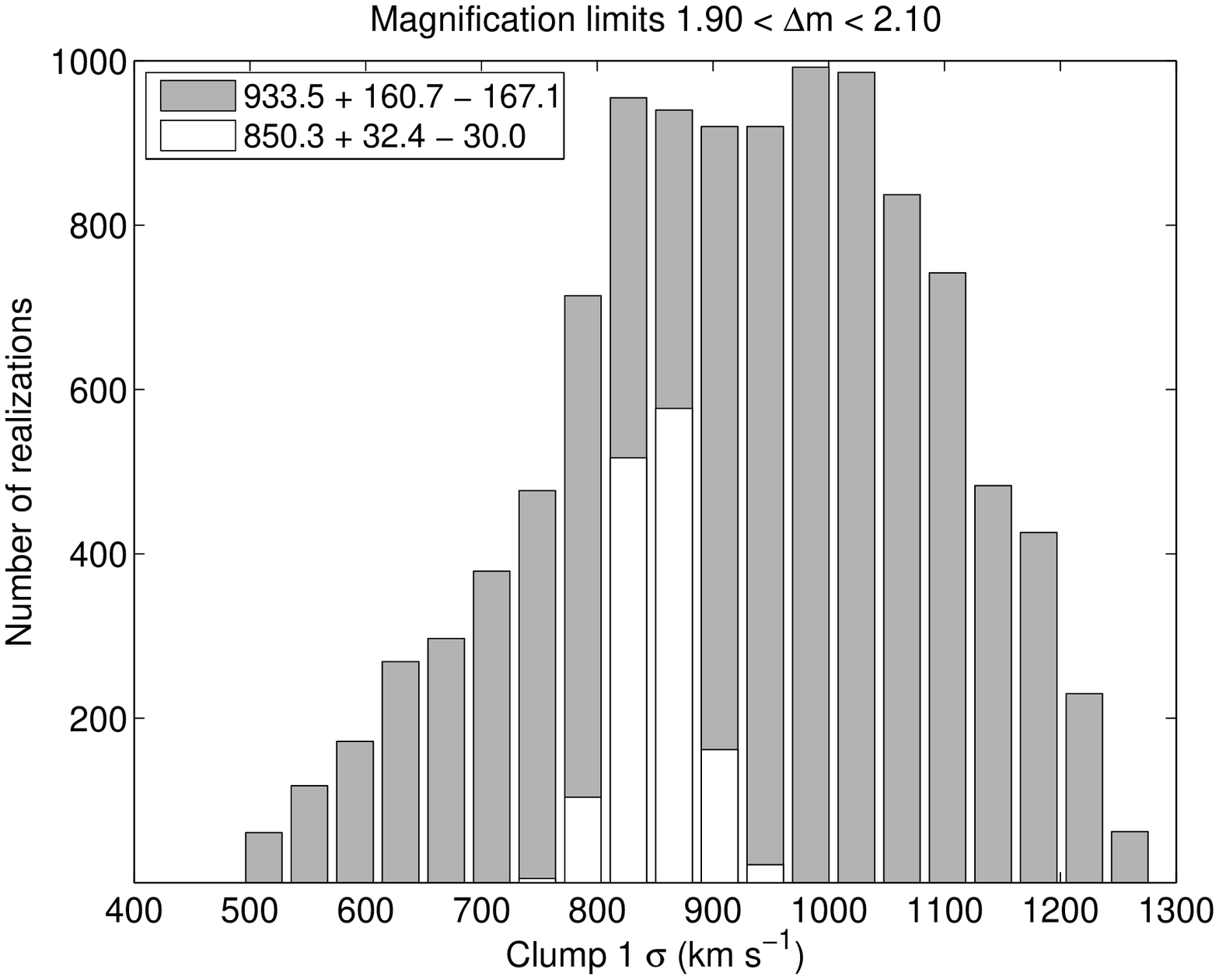,width=0.48\linewidth,clip=} & 
\epsfig{file=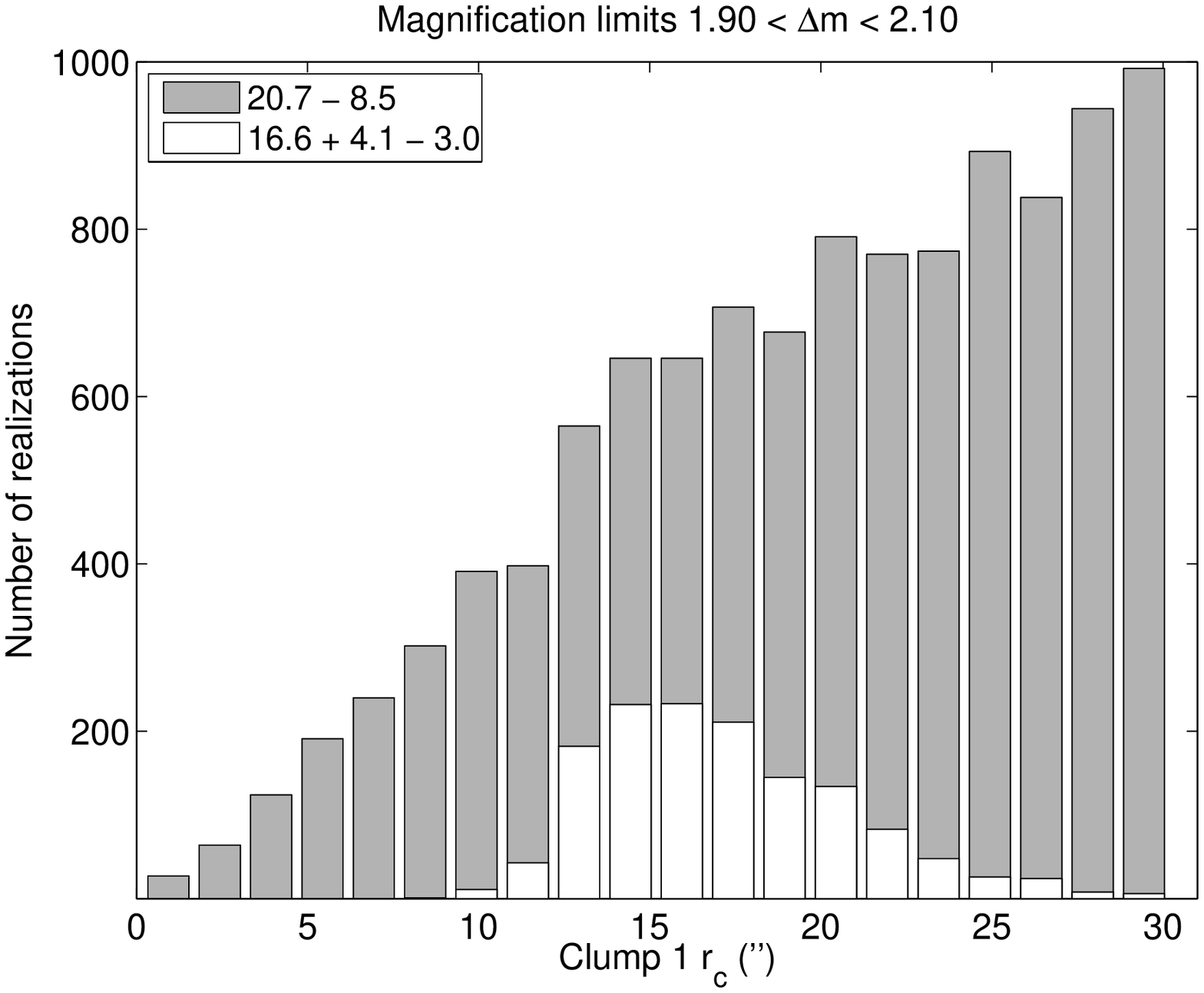,width=0.48\linewidth,clip=}  \\
\end{tabular}

\end{figure*}

In Figure~\ref{fig:2204cuts}, we show the constraints on these
parameters from the magnification of a hypothetical SN~Ia with $z = 1$ 
close to the position of the known lensed image 1.1 measured with 
an uncertainty of $\pm$ 0.1 mag.
The values of the correlation coefficients for $\sigma$ and $r_{\rm core}$ of 
the main clump (Clump 1) are 0.90 and 0.79, respectively. In
this case, the constraints on the velocity dispersion for
the main lensing potential improve from $\sim 17\%$ to $\sim 3.5\%$,
while the corresponding constraint for the core radius goes
from essentially unconstrained (with a forced upper limit from
observations) to $\sim 25\%$. Thus, observations of SNe~Ia
behind galaxy clusters promise to be a powerful tool in constraining
their overall lens potential and thereby the properties of the dark
matter component.

\section{Measuring the Hubble constant}\label{sec:hubble}

It has earlier been suggested to use time delays from multiply-lensed sources to measure the Hubble constant \citep[e.g.,][]{Refsdal64}. For a strong lens system, the arrival time of an image at angular position $\vec{\theta}$ relative to the unlensed case for a corresponding source position $\vec{\beta}$ is given by
\begin{equation}
  \label{eq:tdel}
  \Delta t = \frac{1+z_{\rm L}}{c}\frac{D_{\rm L} D_{\rm S}}{D_{\rm LS}}\left[\frac{( \vec{\theta} - \vec{\beta})^2}{2}-\psi(\vec{\theta})\right],
\end{equation}
where $z_{\rm L}$ is the redshift of the lens, $D_{\rm L}$, $D_{\rm S}$, and 
$D_{\rm LS}$
are, respectively, the angular diameter distances to the lens, to the
source, and from the lens to the source, and $\psi(\vec{\theta})$ is
the lens potential.

Since the ratio of the angular diameter distances scales inversely
with the Hubble constant, $D \equiv \frac{D_{\rm L} D_{\rm S}}{D_{\rm LS}} \propto
H_0^{-1}$. By modeling the lens potential, $\psi(\vec{\theta})$, and
the source position, $\vec{\beta}$, one can use time delay
measurements from strongly lensed systems to estimate the Hubble
constant.

There is also a dependence on $D$ from other cosmological parameters,
such as $\Omega_{\rm M}$ and $\Omega_\Lambda$, although this dependence is much
weaker than that on the Hubble constant 
\citep[e.g.,][]{BoltonBurles03,Coe09,Suyu10}.

Due to the transient nature of SNe, the observed time delay from SN
light curves can be determined with high precision, typically on the
order of less than a few days. To improve the constraints on $H_0$ it
is thus important to be able to narrow down the errors of the lens
potential. Here we investigate the constraints on $H_0$ possible when
observing a SN in one of the known multiply-imaged background galaxies
of A1689 with predicted time delays below 5 years and spectroscopic
redshifts. Summing up our SN rate estimates (given in
Table~\ref{tab:spec_delays}), we expect an average total of 3 SNe
exploding in those galaxies during a survey time of 5 years. The
effective number of strongly lensed SNe which are observable, will
depend on the survey properties. For a five year monthly survey using
the HST/WFC3, we would expect to detect approximately 1 strongly
lensed SN in these galaxies. However, one should keep in mind that
there are 10 more known strong lensing systems behind A1689,
consisting of 30 images, which have not been included in this study
since they have photometric redshifts only. Furthermore, as
observations of this massive cluster continue, additional multiply
lensed systems might be discovered \citep{Coe09}.

\begin{figure}
\caption{Predicted time delays in days and percentual one sigma errors 
for all the image pairs of multiply imaged background galaxies with
spectroscopic redshift and time delay smaller than five years behind
A1689. The color of the markers indicates the redshift of the
systems.}
\label{fig:delay_spread}
\includegraphics[width=9cm]{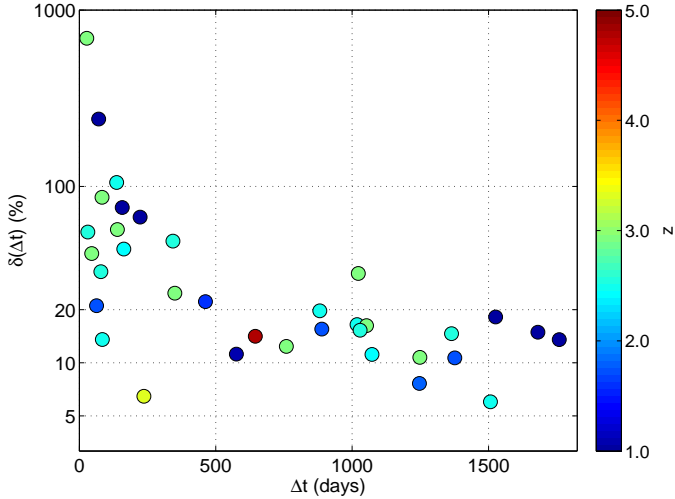}

\end{figure}

From the different realizations of the cluster modeling, we get a
prediction with error estimates for the time delays between image
pairs in the known multiply lensed background galaxies. These errors
depend on the positions of the image pairs. As can be seen in
Fig.~\ref{fig:delay_spread}, for more than half of the image pairs
with predicted time delays below five years and spectroscopic
redshift, the corresponding errors are below 20\%. For several systems
the time delay precision from the model lies around 10\% and
below. Eventually combining constraints from several strongly lensed
SNe, would allow the value of the Hubble constant to be determined
with high precision. Thus, nominally, to match the current $\sim 3\%$ 
accuracy of the local measurements of $H_0$ \citep{Riess11},
 about ten massive clusters should be monitored for five years. Although
it may seem like a rather large time investment in telescope time
to match a result already at hand, we 
emphasize the importance of a non-local test of the universal
Hubble scale, a key test in cosmology.

In the case of a lensed SN~Ia, we can use the information on the
absolute magnification to rule out realizations that predict
magnifications outside the range allowed by observations. With this
extra information, we might be able to further constrain the time
delay error from the lens model. Therefore, we check for correlations
between the predicted magnification of the images in a system and the
time delay between the images.
While many systems show no or only weak correlations between
magnifications and time delay, there are a few more promising systems.

\begin{figure}
\caption{Constraints on the time delay between images 19.3 and 19.4 at 
$z$ = 2.60 behind A1689 using the magnification of both images. In the
shaded histogram the time delay distribution from the complete Monte
Carlo Markov Chain is shown. The unfilled histogram shows the
distribution after the cuts due to magnification constraints. Errors
are given as one sigma.}  
\label{fig:cuts_dmdt} 
\includegraphics[width=9cm]{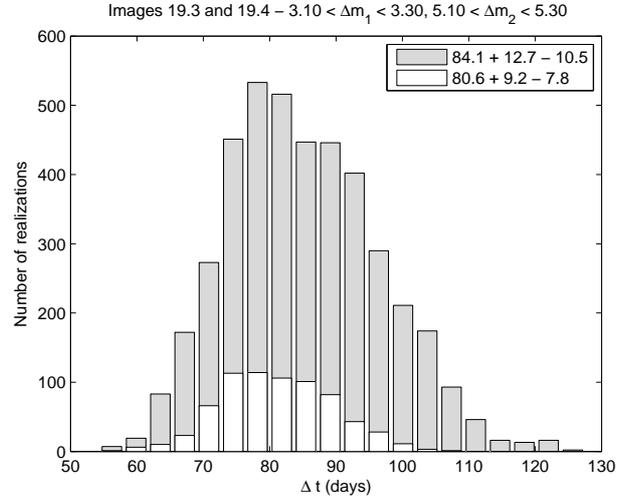}
\end{figure}

Figure~\ref{fig:cuts_dmdt} shows an example for the cuts in time delay
error which can be made using magnification constraints from a lensed
SN~Ia in background galaxy 19. In this case, the errors on the time
delay between images $19.3$ and $19.4$ can be reduced from $\approx
^{+15.1}_{-12.5}$ to $\approx ^{+11.4}_{-9.7}$ \%.

As discussed in Sect.~\ref{sec:modelconstr}, the additional model
constraints from a measurement of the magnification by using the
standard candle nature of SNe~Ia are only modest for a well
constrained cluster as A1689. However, such a constraint would prove
much more valuable in the case of a time delay measured in a less
constrained cluster as it could break the degeneracy between the mass
profile and the Hubble constant \citep{Oguri03}.


\section{Discussion and Conclusions}\label{sec:dis}

A SN~Ia exploding in any of the background galaxies of a cluster will
give an absolute measure of the magnification at that point and can
therefore potentially be used to constrain the mass modeling of the
cluster. For cluster A1689, which we have been focusing on in this
study, the cluster potential is already very well modelled. 
However, such an additional constraint from one or more
lensed SNe could help to improve the cluster model further by reducing
parameter uncertainties. For less constrained galaxy clusters, this
technique proves to be even more interesting. Using the magnification
information from a single SN~Ia behind cluster A2204 can be a powerful
tool to narrow down the model parameters of the dark matter halo
component.

Since SNe are point-like sources for a limited period, microlensing
from stars in a cluster galaxy close to the line of sight might in
principle have an important effect on the magnification and thus
affect the possible constraints on the large scale mass model from SNe~Ia. 
However, it has been concluded \citep{Oguri03} that this effect should only 
cause small deviations. Microlensing events caused by
massive compact halo objects (MACHOs) in the intracluster medium might
unambiguously be identified in SN light curves. Thus any modulation of
the light curve shape, or lack thereof, can put limits on the mass
fraction of the cluster mass in the form of $10^{-7} < M/M_{\odot} <
10^{-4}$ MACHOs \citep{Kolatt98}.

In the case of a SN exploding in one of the known multiply lensed
background galaxies, there is a good chance of observing multiple images
of the SN and determine the observed time delay with high
precision. The intrinsically brighter SNe, e.g. SNe~Ia and SNe IIn,
should in principle be detectable in essentially all of the
systems. However, for SNe IIP, which are expected to be the most common SN type,
the magnification is only strong enough for them to become observable
in about half of the cases.

Due to the well constrained mass model for galaxy cluster A1689, SN
time delays can be used as an independent measure on the
Hubble parameter at high redshift. In case of a SN~Ia, the
magnification might also be used as a constraint to further improve
the error in the time delay predicted by the lensing model. Another
advantage with using a cluster with a well constrained mass model is
the possibility of predicting subsequent images, if a SN
behind the cluster is observed.

It is quite exciting that CLASH, a cluster monitoring multi-cycle
  program is currently been pursued \citep{Postman11}. However, the
  strategy in the CLASH program is tuned for finding and accurately
  studying SNe~Ia in the parallel fields, i.e., in the low magnification
  region. Much of the observations of the cluster cores are done in
  optical and UV filters, where high-$z$ SNe would not be detectable.
  Furthermore, the cluster fields are only monitored for 2-4 months
  making the search for multiple images practically impossible,
  especially considering the time dilation $(1+z)=$2-4.5 in the lensed
  SN lightcurves.

Corroborating the value of $H_0$ found in the local universe could
provide crucial support to our cosmological picture. On the other
hand, a discrepancy would falsify the accepted scenario and shed new
light into the dark matter and dark energy puzzles.  Since strongly
lensed SNe involve both angular diameter and luminosity distances
($d_{\rm A}$ and $d_{\rm L}$), such systems provide an unique chance to probe the
distance reciprocity relation, $d_{\rm L}=(1+z)^2d_{\rm A}$.  Any violation of
this relation would indicate either the existence of unaccounted
astrophysical systematic effect affecting one of the distance measure but not
the other, or that the Universe is not described by the standard
cosmological model. Specifically, a violation could indicate that the
Universe is described by a non-metric theory of gravity or a theory in
which light does not travel on unique null geodesics. Current tests of
$d_{\rm L}/d_{\rm A}$ are restricted to $z<1$ \cite[e.g.,][]{Lampeitl10} and have
$>20 \%$ uncertainties.


\begin{acknowledgements}

EM would like to thank the Swedish Research Council for financial support. 
The Dark Cosmology Centre is funded by the Danish National Research Foundation.
\end{acknowledgements}


\end{document}